\documentclass[%
 reprint,
 amsmath,amssymb,
 aps,
aps,prl,twocolumn,superscriptaddress,floatfix,longbibliography,showpacs]{revtex4-1}

\usepackage{graphicx}
\usepackage{dcolumn}
\usepackage{bm}
\usepackage{xcolor}
\usepackage{capt-of}
\usepackage{soul}

\begin{document}

\preprint{APS/123-QED}

\title{Weyl nodal ring states and Landau quantization with very large magnetoresistance in square-net magnet EuGa$_4$}

\author{Shiming~Lei}
\thanks{These authors contributed equally to this work}
\email{sl160@rice.edu}
\affiliation{Department of Physics and Astronomy, Rice University, Houston, TX, 77005 USA}%

\author{Kevin Allen}
\thanks{These authors contributed equally to this work}
\affiliation{Department of Physics and Astronomy, Rice University, Houston, TX, 77005 USA}


\author{Jianwei Huang}
\affiliation{Department of Physics and Astronomy, Rice University, Houston, TX, 77005 USA}

\author{Jaime M. Moya}
\affiliation{Department of Physics and Astronomy, Rice University, Houston, TX, 77005 USA}
\affiliation{Applied Physics Graduate Program, Rice University, Houston, TX, 77005 USA}

\author{\textcolor{black}{Tsz Chun Wu}}
\affiliation{Department of Physics and Astronomy, Rice University, Houston, TX, 77005 USA}

\author{Brian Casas}
\affiliation{National High Magnetic Field Laboratory, Tallahase, Florida 32310, USA}

\author{Yichen Zhang}
\affiliation{Department of Physics and Astronomy, Rice University, Houston, TX, 77005 USA}

\author{\textcolor{black}{Ji Seop Oh}}
\affiliation{Department of Physics and Astronomy, Rice University, Houston, TX, 77005 USA}
\affiliation{Department of Physics, University of California, Berkeley, California 94720, USA}

\author{Makoto Hashimoto}
\affiliation{Stanford Synchrotron Radiation Lightsource, SLAC National Accelerator Laboratory, Menlo Park, California 94025, USA}

\author{Donghui Lu}
\affiliation{Stanford Synchrotron Radiation Lightsource, SLAC National Accelerator Laboratory, Menlo Park, California 94025, USA}

\author{Jonathan Denlinger}
\affiliation{Advanced Light Source, Lawrence Berkeley National Laboratory, Berkeley, California 94720, USA}

\author{\textcolor{black}{Chris Jozwiak}}
\affiliation{Advanced Light Source, Lawrence Berkeley National Laboratory, Berkeley, California 94720, USA}

\author{\textcolor{black}{Aaron Bostwick}}
\affiliation{Advanced Light Source, Lawrence Berkeley National Laboratory, Berkeley, California 94720, USA}

\author{\textcolor{black}{Eli Rotenberg}}
\affiliation{Advanced Light Source, Lawrence Berkeley National Laboratory, Berkeley, California 94720, USA}

\author{Luis Balicas}
\affiliation{National High Magnetic Field Laboratory, Tallahase, Florida 32310, USA}
\affiliation{Department of Physics, Florida State University, Tallahassee, Florida 32306, USA}

\author{\textcolor{black}{Robert Birgeneau}}
\affiliation{Department of Physics, University of California, Berkeley, California 94720, USA}
\affiliation{Materials Science Division, Lawrence Berkeley National Laboratory, Berkeley, California 94720, USA}

\author{\textcolor{black}{Matthew S. Foster}}
\affiliation{Department of Physics and Astronomy, Rice University, Houston, TX, 77005 USA}
\affiliation{Rice Center for Quantum Materials, Rice University, Houston, Texas 77005, USA}

\author{Ming Yi}
\affiliation{Department of Physics and Astronomy, Rice University, Houston, TX, 77005 USA}

\author{Yan Sun}
\email{sunyan@imr.ac.cn}
\affiliation{Shenyang National Laboratory for Materials Science,Institute of Metal Research, Chinese Academy of Sciences}

\author{Emilia Morosan}
\email{em11@rice.edu}
\affiliation{Department of Physics and Astronomy, Rice University, Houston, TX, 77005 USA}



\date{\today}

\begin{abstract}
\textbf{Magnetic topological semimetals (TSMs) allow for an effective control of the topological electronic states by tuning the spin configuration, and therefore are promising materials for next-generation electronic and spintronic applications. Of magnetic TSMs, Weyl nodal-line (NL) semimetals likely have the most tunability,
and yet they are the least experimentally studied so far due to the scarcity of material candidates. Here, using a combination of angle-resolved photoemission spectroscopy and quantum oscillation measurements, together with density functional theory calculations, we identify the square-net compound EuGa$_4$ as a new magnetic Weyl nodal ring (NR) semimetal, in which the line nodes form closed rings in the vicinity of the Fermi level. Remarkably, the Weyl NR states show distinct Landau quantization with clear spin splitting upon application of a magnetic field.
At 2 K in a field of 14 T, the transverse magnetoresistance of EuGa$_4$ exceeds 200,000\%, which is more than two orders of magnitude larger than that of other known magnetic TSMs. High field magnetoresistance measurements indicate no saturation up to 40 T. \textcolor{black}{Our theoretical model indicates that the nonsaturating MR naturally arises as a consequence of the Weyl NR state.}
Our work thus point to the realization of Weyl NR states in square-net magnetic materials, and opens new avenues for the design of magnetic TSMs with very large magnetoresistance.}
\end{abstract}

\maketitle

Magnetic topological semimetals (TSMs) that are characterized by linear band crossings in momentum space have been established as hosts to many emergent properties, such as Fermi arc surface states \cite{wan2011topological}, the chiral anomaly \cite{burkov2014chiral,hirschberger2016chiral}, large anomalous Hall effect (AHE) \cite{burkov2014anomalous,manna2018colossal,liu2018giant,wang2018large} and drumhead surface states \cite{burkov2011topological,belopolski2019discovery}. Compared to their nonmagnetic counterparts, magnetic TSMs provide a unique opportunity to tune their electronic structure and, consequently, the band topology by manipulating the spin configuration, thus providing an important materials platform for the design of topological electronic and spintronic devices \cite{vsmejkal2018topological,hasan2021weyl, bernevig2022progress}. 

For magnetic TSMs, the band crossings can result in isolated points or lines, giving rise to Weyl points or Weyl nodal line (NL) states, respectively. In principle, 
the formation of the former states requires only the lattice translation symmetry, while the latter demands additional symmetries such as a mirror reflection \cite{chiu2014classification}. When the mirror reflection is destroyed, for example, by rotating the magnetic moments under an applied magnetic field, the Weyl NLs become gapped and Weyl point states emerge \cite{nie2019topological, cano2017chiral}. 

Although there has been great progress in theoretical studies of Weyl NLs in magnetic TSMs \cite{burkov2011topological,phillips2014tunable,belopolski2019discovery,chang2016room, fang2016topological, chang2017topological, jin2017ferromagnetic,nie2019topological}, their experimental realization is rather limited, especially in the presence of spin-orbit coupling (SOC) \cite{hasan2021weyl,bernevig2022progress}. For example, in Fe$_3$GeTe$_2$ the NL states are gapped by SOC, although the gap is small at certain locations in  the momentum space \cite{kim2018large}. Thus far, only the Co-based Heusler alloys Co$_2$MnZ (Z = Ga and Al) \cite{belopolski2019discovery,li2020giant} have been experimentally identified as magnetic Weyl NL semimetals, and only Co$_2$MnGa has gained a good understanding on the electronic structure through spectroscopy measurements \cite{belopolski2019discovery}. Nevertheless,
magnetotransport properties in magnetic Weyl NL semimetals, particularly in the Landau quantized regime, where $\mu B>1$ ($\mu$ is the carrier mobility and \textit{B} is the applied magnetic field) \cite{yang2022quantum, hu2008classical}, are largely unexplored. 
It is imperative to experimentally identify new magnetic Weyl NL candidates, ideally with high carrier mobility, Weyl NL states close to the Fermi level $E_F$, and small energy variation, to maximize their effects on the electronic properties \cite{fang2015topological,bian2016topological,deng2019nodal}.

Here, we report the discovery of Weyl nodal ring (NR, or closed-loop NL) states near E$_F$ in the magnetic square-net EuGa$_4$ in the presence of mirror symmetry protection. Using angle resolved photoemission spectroscopy (ARPES) and quantum oscillation (QO) measurements, we probe the electronic structures of EuGa$_4$ both in the paramagnetic and spin-polarized (SP) states. The good agreement between experimental and density functional theory (DFT) calculation results provides strong evidence for the existence of Weyl NR states with low dispersion along the ring near $E_F$. The quantum mobility is among the highest of all known magnetic TSMs. Associated with the Weyl NR states, we report very large, non-saturating transverse magnetoresistance (MR) up to the Landau quantized regime, exceeding 200,000 \% at $T = 2\,$K and $\mu_0 H = 14\,$T. This value is more than two orders of magnitude higher than that of other known magnetic TSMs, and comparable even with the higher values in nonmagnetic TSMs. \textcolor{black}{Our magnetotransport theory emphasizes the important role of Weyl nodal rings in the non-saturating MR behavior.}

\section{\label{sec:level1}Results\protect\\}

\begin{figure*}[ht]
\centering
\includegraphics[width=0.9\textwidth]{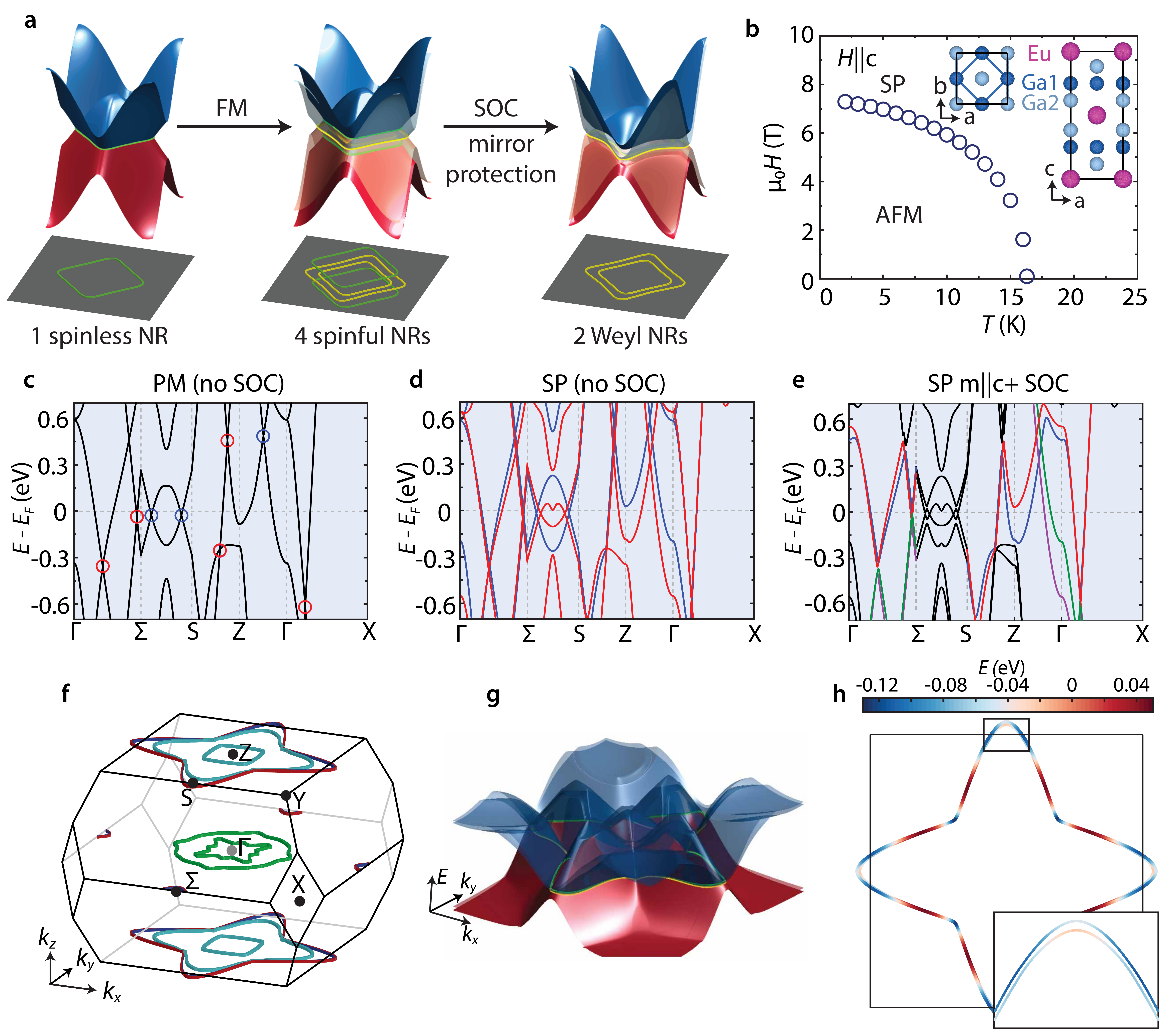}
\caption{\mbox{EuGa$_{4}$ as a candidate to host Weyl NR states}. \textbf{a} The proposed mechanism to create Weyl NR states in square-net magnetic materials: one spinless NR evolves into four spinful NRs and eventually two symmetry-protected Weyl NRs. FM, ferromagnetic; SOC, spin-orbit coupling. \textbf{b}, Magnetic phase diagram ($H-T$) for \mbox{EuGa$_{4}$}. SP, spin-polarized phase; AFM, antiferromagnetic phase. Inset shows the top and side views of the \mbox{EuGa$_{4}$} crystal structure. The empty circle symbols mark the magnetic phase boundary determined by magnetization measurements, see Supplementary Fig. S1 for the full \textit{M}(\textit{H}) data. \textbf{c-e}, Band structures of \mbox{EuGa$_{4}$} in the PM phase without SOC, SP phase without SOC, and SP phase with magnetic moment along c-axis with SOC, respectively. In (\textbf{d}), blue and red indicates two sets of spin-split bands. In (\textbf{e}), the bands that host protected crossings are colored. \textbf{f}, 3D view of the Weyl NRs from DFT calculations. Note that small parts of the red/blue NRs near S on the $k_z=\pm 2\pi/c$ planes extend outside of the BZ. Symmetry operations fold these extended segments back to the $k_z=0$ plane of the BZ. \textbf{g}, Energy surface of the bands that form the red/blue NRs. \textbf{h}, Top view of the red/blue Weyl NRs, with the color indicating the energy. Inset in (\textbf{h}) shows the zoom-in view of the NR pair. The legend is shown on the top.
}
\label{Fig1_theory}
\end{figure*}

\noindent\textbf{Mechanism for Weyl NRs formation in a square lattice.} Weyl NR states in a square lattice emerge as a result of spin degeneracy breaking and SOC, with mirror symmetry protection. Figure \ref{Fig1_theory}a illustrates this mechanism. Without SOC, square-net compounds with conduction bands derived from $p_x/p_y$ orbitals 
serve as a platform to host \textit{spinless} four-fold degenerate diamond-shaped NRs in the mirror invariant plane (left, Fig. \ref{Fig1_theory}a) \cite{lee2013anisotropic,schoop2016dirac,klemenz2019topological,lei2021band}. When ferromagnetism (FM) is introduced (middle, Fig. \ref{Fig1_theory}a), the spin degeneracy is lifted, resulting in four \textit{spinful} NRs, with each NR two-fold degenerate. Finally, when SOC is turned on, only a subset of these spinful NRs survives, depending on the orientation of the magnetic moment $m$. When  $m$ is perpendicular to the mirror plane, the mirror symmetry is preserved. Therefore, one pair of NRs from bands with opposite mirror eigenvalues is protected, while the other pair of NRs with the same mirror eigenvalue is suppressed by opening band gaps. Since the spinless NRs in square-net materials typically have a small energy dispersion \cite{lee2013anisotropic,klemenz2019topological,lei2021band}, this mechanism offers an opportunity to create low-dispersion Weyl NR states.

EuGa$_{4}$, which crystallizes in the BaAl$_{4}$-type structure  (space group $I4/mmm$) \cite{nakamura2013magnetic, nakamura2015transport} with Ga sublattice forming layered square nets (inset, Fig. \ref{Fig1_theory}b), proves suitable for realizing the Weyl NR states following this mechanism. Its magnetic phase diagram is shown in Fig. \ref{Fig1_theory}b. When $H=0$ below $T_\text{N}=16.3\,$ K, EuGa$_4$ is an A-type antiferromagnet (AFM), with the Eu moments parallel to the $a$ axis \cite{kawasaki2016magnetic}. When $H\parallel c$ is applied, the moments rotate towards the field direction until a phase transition to the spin-polarized (SP) state (or field induced FM state \cite{nakamura2013magnetic}).

In the paramagnetic (PM) state above $T_\text{N}$, there are three mirror reflection symmetries for the EuGa$_4$ lattice: m$_z$, m$_x$ (or m$_y$), and m$_{xy}$, where the mirror planes are perpendicular to the $z$, $x$ (or $y$), and the in-plane diagonal crystallographic axis, respectively. When the moments are ordered, at least two of these three mirror reflections are destroyed, depending on the specific magnetic configuration. When the magnetic moments are along the c axis ($m\parallel c$), the Eu layers act as the m$_z$ mirror planes, which allows the formation of Weyl NR states. \textcolor{black}{In Supplementary Sec. 3, we also provide an extended discussion on the mechanism of Weyl NR states in EuGa$_4$ compared to Dirac/Weyl point states in the broad family of square-net topological semimetals.} 

The band structure of EuGa$_4$ from DFT calculations in the PM state, the SP state without SOC, and the SP state with SOC are shown in Figs. \ref{Fig1_theory}c-e, respectively. In the PM state, the bands show multiple crossings, with the corresponding nodes divided into two groups, on (red circles) or off (blue circles) the mirror invariant planes at $k_z=0$ and $k_z=\pm 2\pi/c$ (Fig.~\ref{Fig1_theory}c). 
In the three-dimensional (3D) $k$ space, these nodes, except the one along $\Gamma-\text{Z}$, extend to form lines (\textcolor{black}{Supplementary Fig. S2}). In particular, the NLs on the $k_z=0$ and $k_z=\pm 2\pi/c$ planes exhibit NR geometry. When the spin is fully polarized in the Eu sublattice without SOC, two sets of spin-split bands form (Fig. \ref{Fig1_theory}d). When $m\parallel c$ with SOC, only the crossings from bands with opposite mirror eigenvalues are retained (Fig. \ref{Fig1_theory}e), resulting in the formation of Weyl NRs, as shown in Fig. \ref{Fig1_theory}f. Depending on their band origins, these Weyl NRs can be categorized into three groups: the ones on the $k_z=0$ plane (green), $k_z=\pm 2\pi/c$ planes (red/blue pair), and $k_z=\pm 2\pi/c$ planes (cyan). 
In particular, the red/blue NRs are found to sit very close to $E_F$  with small energy variation of 0.18eV, although they span the whole $k_z=\pm 2\pi/c$ planes of the Brillouin zone (BZ) (Fig.\,\ref{Fig1_theory}f-h). 

To experimentally validate the existence of the Weyl NR states, we provide below ARPES and QO measurements, which allow us to: 1) identify the spinless NR states in the PM state; and 2) determine the band splittings of these NRs in the SP state. When two pairs of spin-split bands cross in the mirror invariant plane, Weyl NR states are guaranteed. 

\begin{figure*}[ht]
\centering
\includegraphics[width=0.78\textwidth]{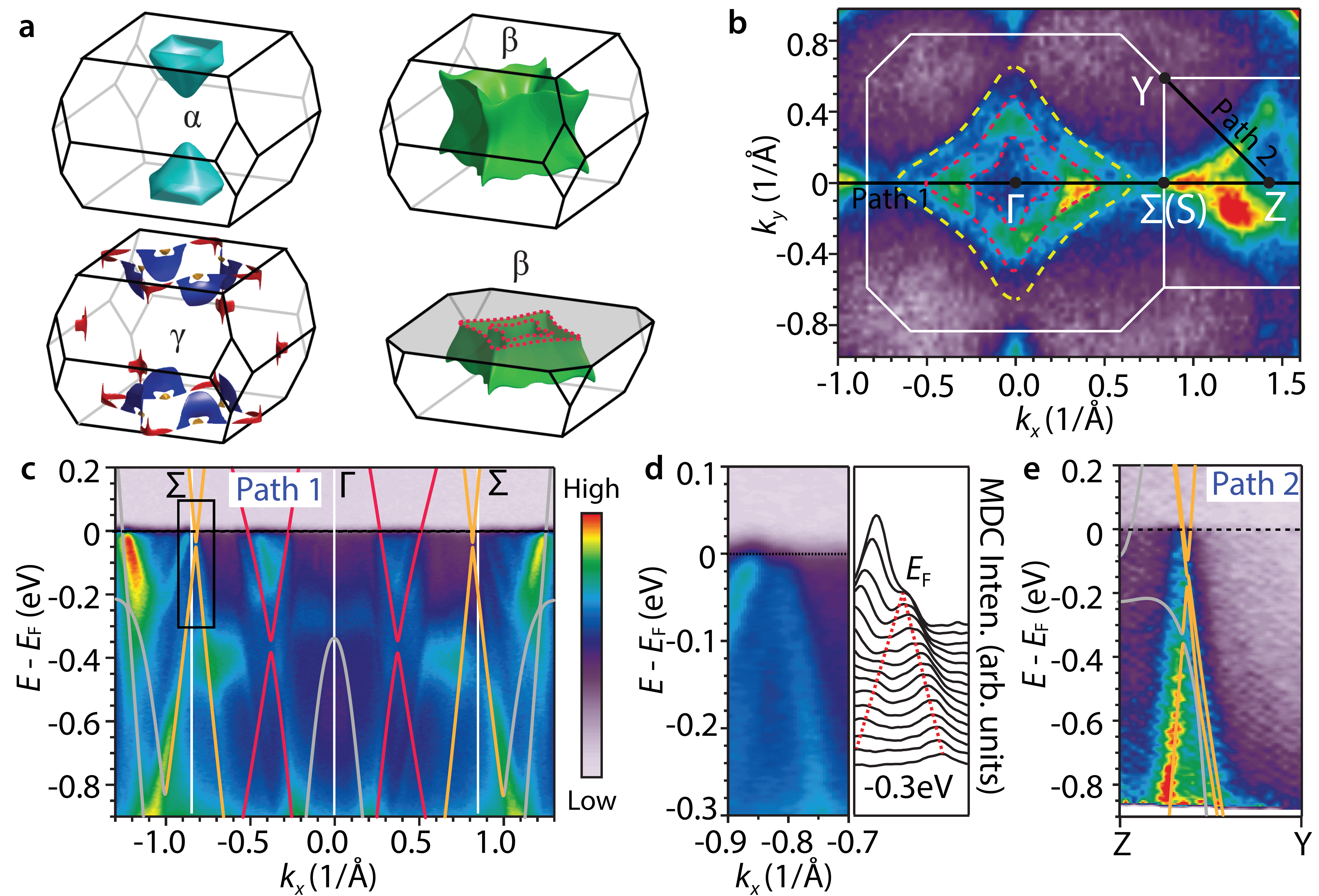}
\caption{Electronic structure of \mbox{EuGa$_{4}$} in the PM phase. \textbf{a}, Three groups of FS pockets: $\alpha$, $\beta$, and $\gamma$, based on DFT calculations. The cross-sectional cut of the $\beta$ pocket at the $k_z=0$ plane is illustrated with dashed red lines. \textbf{b},
ARPES measured FS with $h\nu = 118$\,eV and $T=25\,$K. Two high-symmetry $k-$paths (black lines) are indicated for band dispersion analysis. The white lines mark the BZ boundary. The dashed red lines are the $k_z=0$ cross sections of the $\beta$ pocket, the same as those shown in (\textbf{a}). The dashed yellow lines delineate the boundary within which the effect of $k_z$-broadening is observed, as discussed in the text. \textbf{c}, ARPES band dispersion along path 1 with $h\nu = 120$\,eV. The solid lines are band structures from DFT calculations. Red and orange indicate the bands that form the NR1 and NR2, respectively. 
\textbf{d}, Zoom-in view of the boxed region in (\textbf{c}), with the MDC stacks shown on the right. \textbf{e}, ARPES band dispersion along the $\text{Z}-\text{Y}$ path.
}
\label{Fig2_APRES}
\vspace{-5pt}
\end{figure*}

\vspace{3mm}
\noindent\textbf{ARPES investigation of spinless NR states.} 
As shown in Fig.~\ref{Fig1_theory}c, there are two crossings along the $\Gamma-\Sigma$ path; one is $0.36\,$eV below $E_{F}$ and the other very close to $E_{F}$. These two crossings and the one above $E_F$ on the $k_z=\pm 2\pi/c$ plane extend to form three spinless NRs in the \textit{k} space (denoted as NR1, NR2, and NR3, see  Supplementary Fig. S2a). The Fermi surface (FS) pockets derived from these NR bands are accordingly divided into three groups: $\alpha$, $\beta$, and $\gamma$, as shown in Fig. \ref{Fig2_APRES}a.

Figure \ref{Fig2_APRES}b shows the measured FS cross section of EuGa$_4$ at $25\,$K (PM phase), with a photon energy $h\nu= 118$\,eV, which corresponds to the $k_z \approx 0$ plane  (for photon energy dependent data see \textcolor{black}{Supplementary Fig. S3}). Centered at the $\Gamma$ point, the ARPES data show enhanced intensity within two concentric diamond rings (dashed red curves in Fig. \ref{Fig2_APRES}b), which are exactly the inner and outer $k_z=0$ cross sections of the $\beta$ pocket from DFT calculations (Fig. \ref{Fig2_APRES}a). Outside the outer diamond, finite ARPES intensity, albeit lower than the region in between, persists up to the dashed yellow boundary. The origin of its nonzero ARPES intensity is attributed to $k_z$ broadening, considering the outward warping geometry of the $\beta$ pocket along $k_z$.

To view the band dispersion, we extracted the measured ARPES spectra along two high-symmetry paths, one along $\Sigma-\Gamma-\Sigma\left(\text{S}\right)-\text{Z}$  (Fig. ~\ref{Fig2_APRES}c), and the other along the diagonal $\text{Z}-\text{Y}$ direction  (Fig. ~\ref{Fig2_APRES}e). For comparison, the DFT calculated band structure (lines) is overlaid on top, with red and orange indicating the NR1 and NR2 bands, respectively. Indeed, the nodes of the NR2 sit very close to $E_F$, as is evident from the zoom-in band image and associated momentum distribution curves in Fig. \ref{Fig2_APRES}d.
\textcolor{black}{Our data further show suppressed spectral weight near $E_F$, suggesting the existence of a small gap. This is consistent with SOC induced gap (20meV) at the crossing from DFT calculations.} The ARPES spectra along the $\text{Z}-\text{Y}$ path (Fig. \ref{Fig2_APRES}e) also show clear linear band crossings near $E_F$, supporting the low dispersion feature along the ring for the NR2. As for the NR1, one branch of the bands appears to be clearer than the other (See the band dispersion along $\Gamma-\Sigma$ in Supplementary Fig. S3b), possibly due to the matrix element effect. 

Overall, the ARPES data supports the existence of spinless NR1 and NR2 in the PM phase of EuGa$_4$, as predicted by theory. Particularly, the NR2 is confirmed to sit very close to $E_F$ with small energy variation along the ring. 

\vspace{3mm}
\noindent\textbf{Weyl NR states in the SP state.}
As shown in Fig. \ref{Fig1_theory}f, there are three pairs of Weyl NRs in the SP state of EuGa$_4$. Consequently, there are three groups of FS pockets (\textcolor{black}{Supplementary Fig. S5}), which appear in pairs (one smaller and one larger) due to band splittings, although the shape is similar to that in the PM state (Fig. \ref{Fig2_APRES}a). Quantum oscillations, which are a direct measure of the FS pockets, 
provide quantitative information about the band splitting and the energy of the Weyl NR states. The oscillation frequency $f$ is related to the cross-sectional area $A_k$ of the FS perpendicular to the applied magnetic field via the Onsager relation: $f = (\Phi_0/2\pi^2)A_k$, where $\Phi_0=2.07\times10^{-15}$\,Tm$^2$ is the flux quantum. By rotating the field, the QO frequency picks up an angle dependence, from which a 3D picture of the shape and size of the FS can be constructed. 

\begin{figure*}[ht]
\centering
\includegraphics[width=1\textwidth]{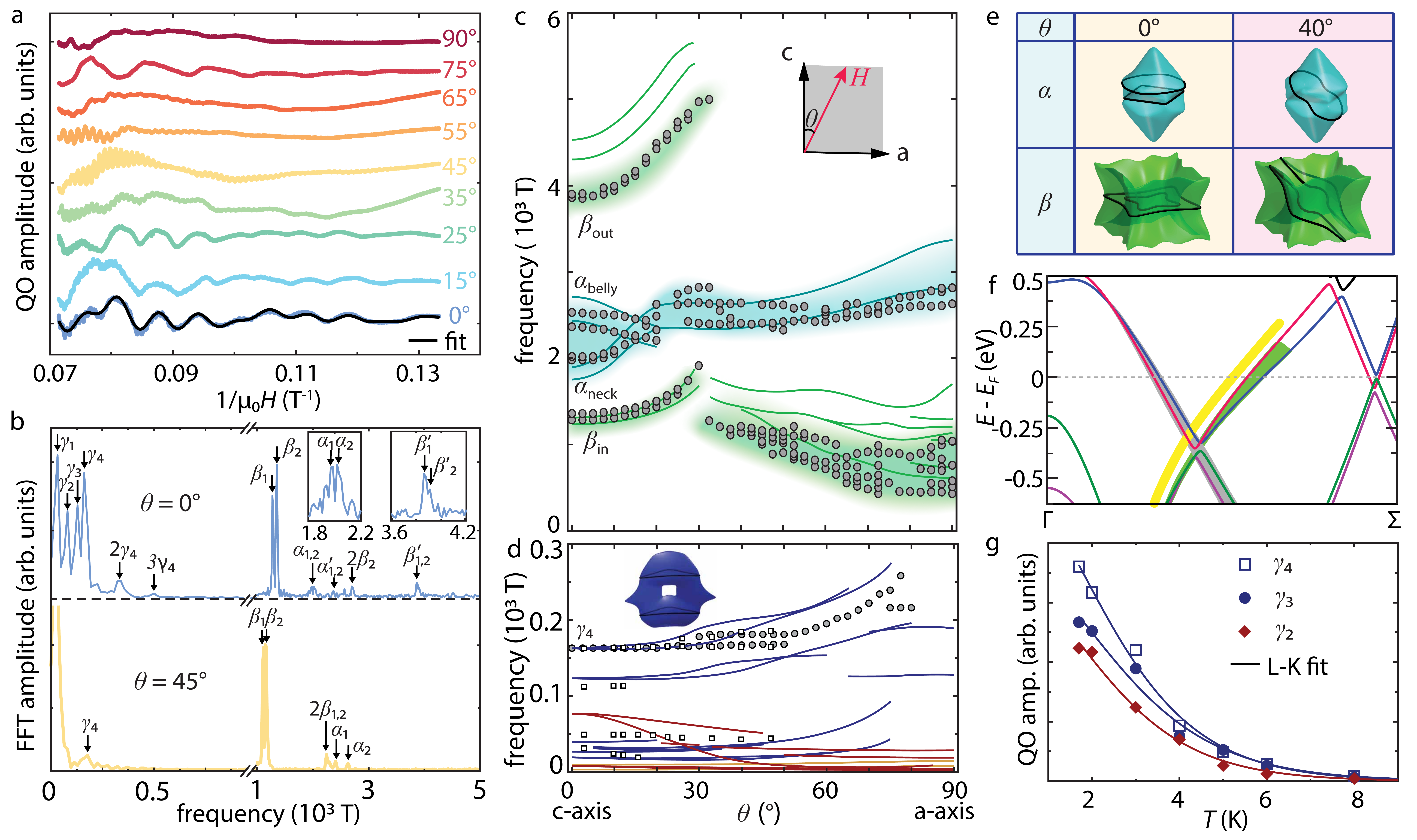}
\caption{Fermi surface geometry of EuGa$_4$ in SP phase from quantum oscillations. \textbf{a}, A series of QO curves with $\theta$ ranging from $0^\circ$ to $90^\circ$. For the QO at $0^\circ$, a L-K fit is shown. \textbf{b}, Two representative FFT spectra of the QOs at $\theta = 0^\circ$ and 45$^\circ$. \textbf{c},\textbf{d} Angle dependent QO frequencies (circles) above and below 300 T, respectively. The shadings act as a guide to the eyes. The theoretical predictions (lines) are shown for comparison. Inset in (\textbf{c}) illustrates the definition of rotation angle, $\theta$. Inset in (\textbf{d}) illustrates the extremal cyclotron orbits associated with the measured $\gamma_4$ frequency at $\theta = 0^\circ$. In Supplementary S6d, we illustrated all the extremal orbits with the $\gamma$ pockets. Note that in (\textbf{d}), the QO frequencies determined from high-field measurement are also shown (square symbols). \textbf{e}, Illustration of the extremal orbits (black lines) of the $\alpha$ and $\beta$ pockets, when $\theta = 0^\circ$ and 40$^\circ$. \textbf{f}, Band structure along $\Gamma-\Sigma$ with the feedback from QO measurements. The energy of the gray shaded bands is accurately described by theory based on the QO measurements,  while that of the green shaded band is underestimated by theory. The actual bands should have a slightly higher energy, as indicated by the yellow shades. \textbf{g}, Temperature dependent QO amplitude for the $\gamma_4$, $\gamma_3$, and $\gamma_2$ oscillations at $\theta=0^\circ$ and their L-K fits (solid lines).
}

\label{Fig3_EuGa4_QO}
\end{figure*}

In Fig. \ref{Fig3_EuGa4_QO}a, we present a series of Shubnikov–de Haas (SdH) oscillations, with the field tilting from $H\parallel c$ ($\theta=0^\circ$) towards $H\parallel a$ ($\theta=90^\circ$). Figure \ref{Fig3_EuGa4_QO}b shows the fast Fourier transform (FFT) analysis of the QOs at two discrete angles: $\theta=0^\circ$  and $45^\circ$. The contour plot of the FFT intensity at all measured angles is shown in the \textcolor{black}{Supplementary Fig. S6} and the extracted QO frequencies are shown as circles in Figs. \ref{Fig3_EuGa4_QO}c,d. 

At high frequencies ($f>300\,$ T, Fig. \ref{Fig3_EuGa4_QO}c), the experimental data show good agreement with the theoretical prediction (colored lines) on the $\alpha$ and $\beta$ pockets, with the colored shading as a guide. Starting at $\theta=0^\circ$, four pairs of QO frequencies (indicated by $\beta_\text{in}$, $\beta_\text{out}$, $\alpha_\text{neck}$, and $\alpha_\text{belly}$) are identified. The two close-lying frequencies within each pair have similar angular dependency, suggesting similar shape of FS and pointing to band splitting as their origin. As $\theta$ gradually increases from $0^\circ$ to $\sim20^\circ$, the $\alpha_\text{neck}$ and $\alpha_\text{belly}$ frequencies merge. By contrast, the $\beta_\text{in}$ and $\beta_\text{out}$ frequencies both increase with $\theta$, until a sudden drop occurs at $\sim30^\circ$. These features suggests a morphological change of extremal cyclotron orbits as the field rotates, which is the key to understanding the shape of probed FS pocket. The $\alpha_\text{neck}$ and $\alpha_\text{belly}$ frequencies at small $\theta$ arise because the $\alpha$ pockets have slight corrugations along the vertical axis, while the $\beta_\text{in}$ and $\beta_\text{out}$ frequencies are associated with the inner and outer cross-sectional areas of the torus-shaped $\beta$ pockets (Fig. \ref{Fig3_EuGa4_QO}e). As $\theta$ increases beyond a certain critical angle ($\theta_\text{c}\sim30^\circ$), the extremal cross section of the $\beta$ pocket undergoes a change from the in-and-out to the sidewise pair.

With the shape of the $\alpha$ and $\beta$ pockets determined, we now evaluate the energy of the bands, with a focus on the QO data at $\theta=0^\circ$ ($H\parallel c$). Figure ~\ref{Fig3_EuGa4_QO}f shows the DFT calculated band structure along the $\Gamma-\Sigma$ path. The bands that give rise to  $\beta_\text{in}$ and $\beta_\text{out}$ oscillations are marked with gray and green shadings, respectively. The excellent match between experiment and theory on the $\beta_\text{in}$ frequencies (Fig.\,\ref{Fig3_EuGa4_QO}c) indicates the accuracy of the $\beta_\text{in}$ bands (gray shading, Fig.~\ref{Fig3_EuGa4_QO}f) from DFT calculations. By comparison, the $\beta_\text{out}$ frequency pair is about $400-600$ T below the theoretical prediction, which means that the actual $\beta_\text{out}$ bands (yellow shading) have slightly higher energy than the theoretical ones (green shading, Fig.~\ref{Fig3_EuGa4_QO}f). Assuming a rigid band shift, the actual band energy is $\sim90-100$ meV higher than the theoretical one. As for the $\alpha_\text{neck}$ and $\alpha_\text{belly}$ frequencies, the experimental ones are slightly higher and lower, respectively, than the theoretical predictions. This result suggests that the extent of neck-and-belly  corrugation of the $\alpha$ pocket is less prominent than predicted by theory. Finally, based on the frequency difference of the $\beta_\text{in}$, $\beta_\text{out}$, $\alpha_\text{neck}$, and $\alpha_\text{belly}$ pairs, the energy of band splittings at $E_F$ are determined to be 45, 10, 17, and 24\,meV, respectively.

We now discuss the QO features of the $\gamma$ pockets. According to the theoretical prediction, they are essentially composed of a series of side-by-side electron and hole pockets (\textcolor{black}{Supplementary Fig. S5c}) along the red/blue Weyl NRs. Since the energy of the nodes is very close to $E_F$, these pockets are all small, giving rise to low-frequency QOs (Fig. \ref{Fig3_EuGa4_QO}d). FFT analysis of the measured QOs at $\theta~=~0^\circ$ reveals four frequency components: $\gamma_1= 30$\,T, $\gamma_2= 77$\,T,  $\gamma_3= 125$\,T, and  $\gamma_4= 163$\,T (Fig. \ref{Fig3_EuGa4_QO}b). A Lifshitz-Kosevich (L-K) fit (Fig.~\ref{Fig3_EuGa4_QO}a) based on these four components reproduces well the measured QO curve. DFT calculations suggest that the blue pocket (inset, Fig. \ref{Fig3_EuGa4_QO}d) has the largest cross-sectional area at $\theta=0^\circ$. As $\theta$ increases, the predicted $f$ remains nearly constant, and gradually bifurcates into two branches, eventually merging into one observable frequency at around $50^\circ$ with weak increase with $\theta$. Such subtle angle dependent behavior is captured by the measured $\gamma_4$ QO frequency, albeit with frequency values slightly smaller than the theoretical ones at high angles, as shown in Fig.~\ref{Fig3_EuGa4_QO}d. Therefore, the $\gamma_4$ frequency is identified as the signature of the blue pocket. Further high-field measurements reveal the existence of three smaller pockets (square symbols in Fig.~\ref{Fig3_EuGa4_QO}d). However, the nature of these pockets is less obvious than the $\gamma_4$ one, and more discussion is included in the Supplementary Secs. 6 and 7.  
Due to the thermal broadening of chemical potential, the QO amplitude decreases with temperature, which provides a way to evaluate the effective mass and quantum mobility
\cite{shoenberg2009magnetic}. The measured QOs at different temperatures are shown in \textcolor{black}{Supplementary Fig. S9}. Based on the L-K fit to the temperature dependent QO amplitude (Fig. \ref{Fig3_EuGa4_QO}g), the effective masses of the $\gamma_4$, $\gamma_3$, and $\gamma_2$ components are:
$m^*\left(\gamma_4\right)=0.74m_e$, $m^*\left(\gamma_3\right)=0.68m_e$, and $m^*\left(\gamma_2\right)=0.76m_e$, where $m_e$ is the electron mass. These values are much higher than those in typical nonmagnetic TSMs, such as Cd$_3$As$_2$ ($0.045m_e$) \cite{liang2015ultrahigh} and NbP ($0.076m_e$) \cite{shekhar2015extremely}, and are also significantly higher than the DFT predictions ($0.02m_e-0.18m_e$) based on the single particle picture (see Supplementary Sec. 8). The mass enhancement reflects the existence of electronic correlation effects in EuGa$_4$. The quantum mobility at $1.7\,$K is estimated to be 830, 1180, and 1630 cm$^2$/Vs for the $\gamma_4$, $\gamma_3$, and $\gamma_2$ components, respectively, among the highest in all known magnetic TSMs. For comparison, the quantum mobility of the magnetic Weyl semimetal Co$_3$Sn$_2$S$_2$ at $1.6\,$K is $106-221\,$ cm$^2$/Vs \cite{ding2021quantum}, which is about one order of magnitude smaller than that in EuGa$_4$.

Overall, with the identification of the spin-split bands for the $\alpha$, $\beta$ and $\gamma$ pockets, our QO data provide strong evidence for the existence of Weyl NR states in EuGa$_4$, as predicted by theory. In particular, the red/blue NRs do cross $E_F$ with small energy variation, giving rise to a series of small pockets, as revealed by the low-frequency QOs. In addition, the QO measurements reveal high quantum mobility. 

\begin{figure*}[ht]
\centering
\includegraphics[width=1\textwidth]{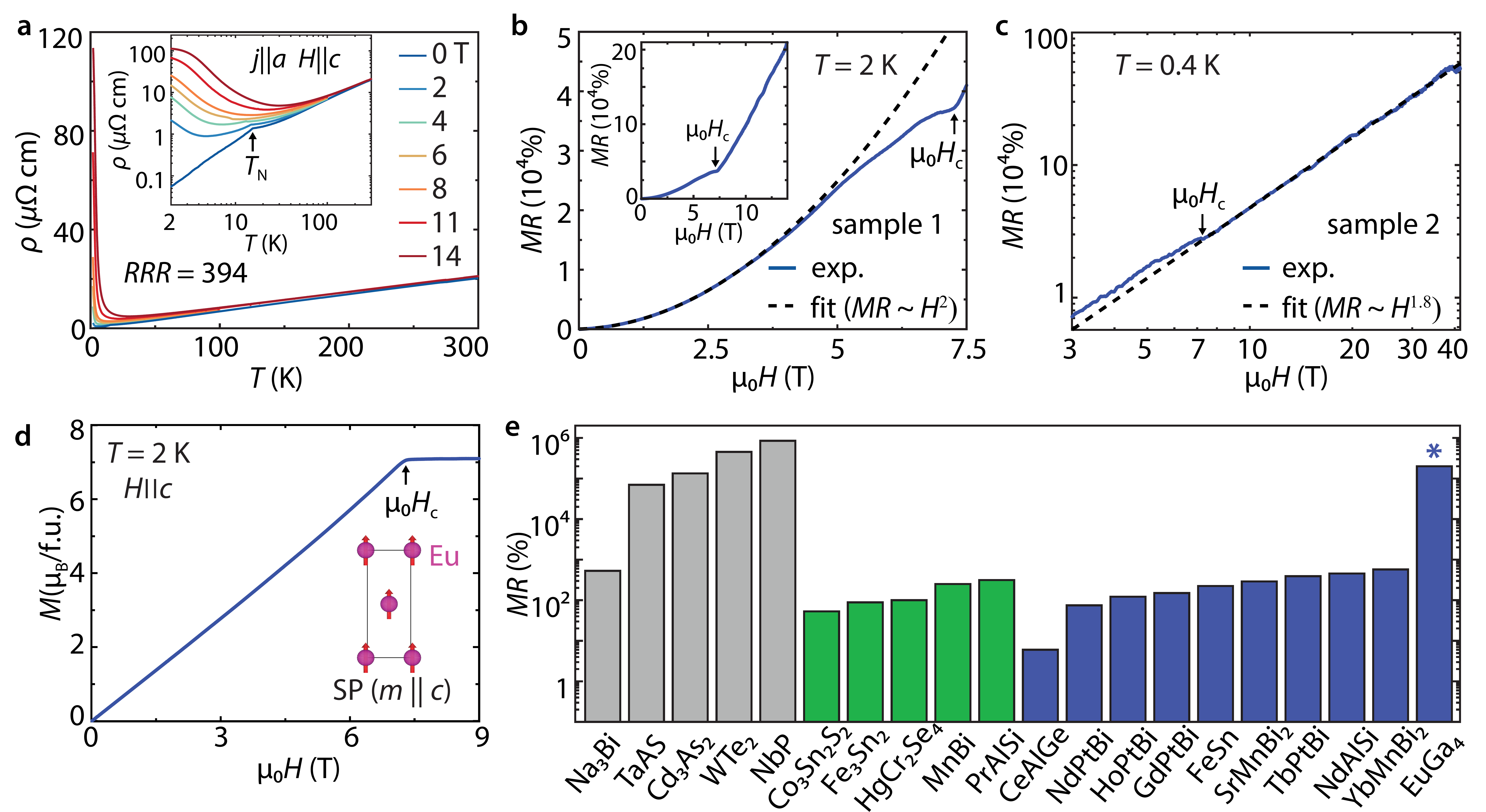}
\caption{Large, non-saturating MR in EuGa$_4$. \textbf{a}, Temperature dependent resistivity at selected fields. Inset shows the same data in a logarithmic scale, highlighting the low-\textit{T} behaviors. The field is applied along the \textit{c}-axis ($H\parallel c$), while the current is along the \textit{a}-axis ($j\parallel a$). \textbf{b,c}, The low- and high-field MR behaviors, respectively. The AFM-to-SP magnetic phase transition is indicated by the arrow. The $H^2$ fit is performed on the MR curve from 0 to $3.5\,$T in (\textbf{b}), while a power function fit is performed above $H_c$ up to $41.5\,$T in (\textbf{c}), with 1.8 as the exponent. Inset in (\textbf{b}) shows the MR curve up to $14\,$T. \textbf{d}, Isothermal magnetization curves ($H\parallel c$) at $T=2\,$K. Fully spin-polarized phase is reached above $\mu_0H_\text{c}=7.4\,$T. Inset illustrates the SP state with the moments on Eu sublattices along the \textit{c}-axis.
\textbf{e}, Comparison of the measured MR in EuGa$_4$ (marked by *) with other known TSMs. The nonmagnetic compounds are colored in gray, while the ferromagnetic and antiferromagnetic ones are colored in green and blue, respectively. See Supplementary Table S1 for the field, temperature and reference information on the compounds in this plot.  
}
\label{Fig4_EuGa4_MR}
\end{figure*}

\vspace{3mm}
\noindent\textbf{Electrical transport properties.} 
The temperature dependent resistivity $\rho\left(T\right)$ of EuGa$_4$ in zero field (Fig. \ref{Fig4_EuGa4_MR}a) reveals a typical metallic behavior, as $\rho$ decreases monotonically with decreasing $T$ down to $2\,$K. Below $T_\text{N}=16.3\,$K, the loss of spin disorder scattering induced a sharp drop, consistent with prior measurements \cite{nakamura2015transport,zhang2021giant}. The high residual resistivity ratio $RRR = \rho\left(300\,\text{K}\right)/\rho\left(2\,\text{K}\right)$ = 394 is indicative of high crystal quality. When $H\parallel c$ is applied, the low-\textit{T} resistivity exhibits an upturn on cooling. Such ``turn-on'' behavior by field suggests large MR response and high transport mobility, which have been seen in several representative nonmagnetic TSMs, such as TaAs \cite{huang2015observation} and NbP \cite{shekhar2015extremely}.

We are interested in the field dependence of the MR response. Qualitatively different field dependence is observed below and above $\mu_0H_{c}=7.4\,$T (Fig.~\ref{Fig4_EuGa4_MR}b,c, $\mu_0$ is the vacuum permeability), which marks the magnetic phase transition at $2\,$K in EuGa$_4$. The MR response in the AFM phase can be well described by a $H^2$ dependence below $\sim3.5\,$ T, and levels off as $\mu_0H$ increases further towards $\mu_0H_{c}$ (Fig. \ref{Fig4_EuGa4_MR}b). This is a typical behavior seen in uncompensated semimetals \cite{pippard1989magnetoresistance}. However, once the system enters the SP phase ($H>H_{c}$), where the Weyl NR states are hosted, the MR shows an abrupt upturn  (Fig. \ref{Fig4_EuGa4_MR}b inset) and continues to increase without any signs of saturation up to $\sim40\,$T (Fig. \ref{Fig4_EuGa4_MR}c and Supplementary Fig. S10).

In AFM metals, field induced spin fluctuations can be a cause for an increase in resistivity. However, as the field continuously increases in the fully SP state, the electron scattering caused by spin fluctuations is increasingly suppressed, which should lead to a resistivity decrease. This appears in quite a few AFM metals, such as MnBi$_2$Te$_4$ \cite{liu2020robust} and EuPtSi \cite{kakihana2018giant}, but is not the case in EuGa$_4$. \textcolor{black}{Furthermore, the carrier compensation mechanism for the nonsaturating MR demands a perfect balance of electron and hole carrier density, $n_e=n_h$; a slight deviation from this condition will lead to a saturating MR at high field \cite{ali2014large}. This is the case in Bi, where the MR deviates from the power law scaling at around 6~T, and reaches full saturation at $\sim 30$~T \cite{fauque2018magnetoresistance}. Given the carrier density in EuGa$_4$ does not meet the carrier compensation condition (see Supplementary Sec. 11 for the estimate of the carrier density), a different mechanism is expected to explain the nonsaturating MR behavior. }


\textcolor{black}{We investigated the magnetotransport properties of a Weyl nodal ring semimetal with both semi-classical and fully quantum mechanical approaches (see Supplementary Sec. 12 for more details on the theoretical modellings and discussions). We find that nonsaturating MR naturally arises in the Weyl nodal ring system without the requirement of perfect electron-and-hole carrier compensation. This unusual behavior benefits from the negligibly small Hall conductivity, which occurs due to the sign reversal of the Fermi velocity across the nodal ring. Notably, our theoretical model also predicts sub-quadratic power-law scaling for MR, which resonates with the experimental observations (Fig.~\ref{Fig4_EuGa4_MR}c). Here we further note that this mechanism is different from the quantum magnetoresistance proposed by Abrikosov \cite{abrikosov1998quantum}, where linear and nonsaturating MR is achieved only when electrons are forced to occupy the lowest Landau level (quantum limit) in a linear-band system. Since most of the conducting carriers come from the large $\alpha-$ and $\beta-$pockets in EuGa$_4$, the Abrikosov mechanism is not expected to play a dominant role in the nonsaturating MR behavior.} 



When comparing the MR in EuGa$_4$ with the values in other known magnetic TSMs (Fig. \ref{Fig4_EuGa4_MR}e), EuGa$_4$ stands out. The MR at $2\,$K and $14\,$T exceeds $2\times10^{5}\%$, which is more than two orders of magnitude larger than in other known magnetic TSMs (green and blue), and even comparable to that in the nonmagnetic ones (grey). As the field increases up to $\sim40\,$T, a non-saturating MR$\sim5\times10^{5}\,$\% is observed in EuGa$_4$ (Fig. \ref{Fig4_EuGa4_MR}c). 

We emphasize that the AFM state of EuGa$_4$ with or without moment canting would fail to provide the required symmetry protection for the existence of Weyl NR states. In the AFM ground state, the two magnetic sublattices ($m\parallel a$) are connected by the joint translation and time-reversal symmetry $\{\text{T}|(1/2, 1/2, 1/2)\}$. Consequently, the spin degeneracy is not lifted and no Weyl NR states are supported in the presence of SOC. In the spin-canted state, the $\{\text{T}|(1/2, 1/2, 1/2)\}$ symmetry is broken along with spin splitting. However, since the spin canting breaks all the mirror symmetries: m$_x$, m$_{xy}$ and m$_z$, no Weyl NRs should exist either. Only when the system reaches the SP state above $H_c$ and the mirror symmetry m$_z$ is recovered, the Weyl NR states appear. Based on this symmetry analysis, we conclude that a field-induced topological phase transition occurs in EuGa$_4$ along with the AFM-SP magnetic transition. The upturn sub-quadratic MR increase is related to this transition.

\section{Conclusions}

We present the magnetic square-net compound EuGa$_4$ as a new host for Weyl NR states. Our combined ARPES and QO measurements provide strong evidence for the existence of Weyl NRs close to the Fermi level in the SP phase, consistent with our DFT predictions. In particular, one pair of Weyl NRs are found to cross $E_F$, with small energy variation of $165-195\,$ meV although it spans the whole plane of the BZ. \textcolor{black}{With high carrier mobility in EuGa$_4$, we reveal clear features of Landau quantization of these NR states. Arguably, the most interesting feature is the qualitatively different field dependent MR behaviors in the AFM and SP phase, where the Weyl NR states are stabilized only in the latter.} While the MR curves in the AFM phase gradually level off, they pick up a fast upturn increase without any sign of saturation up to $\sim40\,$T in the SP phase. These behaviors cannot be attributed to a carrier compensation mechanism. \textcolor{black}{Instead, we developed a theoretical model that naturally explains the nonsaturating MR, highlighting the role of Weyl nodal ring state.}
At $14\,$ T and $2\,$ K, the measured MR exceeds $2~\times10^5~\%$, more than two orders of magnitude larger than in other known magnetic TSMs. Our work thus provide new insight for the design of magnetic materials with large MR.


\section{Methods}
\noindent\textbf{Sample growth and characterization.} Single crystals of EuGa$_{4}$ were grown in an excess of gallium (Ga) via a self-flux technique. Europium (Eu) and Ga were mixed in a ratio of 1:9 then placed in an alumna crucible and evacuated in a quartz ampule. The mix was heated to 900$^{\circ}$C over 2 hours and subsequently slowly cooled over a period of 60 hours down to 700$^{\circ}$C, after which the crystals were separated from the excess liquid flux using a centrifuge. EuGa$_{4}$ forms as plate-like crystals with the biggest surface area corresponding to the crystallographic $a-b$ plane. The largest crystals have lateral sizes up to 5mm. The single crystals were confirmed to have BaAl$_4$ type of structure with powder x-ray diffraction. A Rietveld structural refinement was achieved and fit to the measured intensities. We extracted the structural parameters for EuGa$_{4}$, listed in \textcolor{black}{Supplementary Table S2}, which served as the input for the density functional theory calculations. 

\noindent\textbf{Angle-resolved photoemission spectroscopy experiments.} ARPES experiments were carried out at Beamline 5-2 of the Stanford Synchrotron Radiation Lightsource (SSRL), Beamline 4.0.3, and \textcolor{black}{Beamline 7.0.2 (MAESTRO)} of the Advanced Light Source. 
EuGa$_{4}$ samples were cleaved in-situ to expose the (001) surface in  an ultrahigh vacuum chamber with base pressure $3\times10^{-11}\,$Torr. The ARPES data were acquired within 5 hours after cleaving to minimize the effects of surface degradation. 
The lateral size of the beam is smaller than $50\times50\,\mu$m$^2$. Fermi surfaces and energy-momentum dispersions in Fig. \ref{Fig2_APRES} were collected at 118 and 120 eV, covering the entire Brillouin zone. The photon energy dependent data along $\overline{\Sigma}$-$\overline{\Gamma}$-$\overline{\Sigma}$ path were taken with photon energies ranging from 60 to 180 eV. 

\noindent\textbf{Electrical transport and SdH oscillation measurements.} 
The electrical transport and SdH quantum oscillation experiments were carried out in a standard four-probe geometry in a lab magnetometer, Quantum Design DynaCool system, with field up to $14\,$T. The high-field measurements were performed at the National High Magnetic Field Laboratory at Tallahassee, with fields up to $41.5\,$T. 

The angular dependent QO measurements in the Dynacool system were performed by rotating the sample in the a-c plane from $H\parallel c$ to $H\parallel a$, with the current along b ($j\parallel b$). When $H\parallel c$, we measured QOs at different temperatures to evaluate the cyclotron effective mass. 
The oscillations were obtained after subtracting a polynomial background from the field dependent resistivity data, after which they were analyzed with a fast Fourier transform as a function of inverse field. The angle dependent QO frequencies from the $\alpha$, $\beta$ and $\gamma$ pockets were extracted to compare with theory. To evaluate the cyclotron effective mass ($m^*$) and estimate the quantum lifetime ($\tau_q$) of the $\gamma$ FS pockets, we performed L-K fits to the measurements with four frequency components. Each QO component is described by:

\begin{equation*}
    \Delta \rho \propto \frac{\lambda T}{\sinh (\lambda T)}e^{-\lambda T} \cos \left[ 2\pi \left( \frac{f}{B} - \frac{1}{2} + \beta + \delta \right)\right]
\end{equation*}

where $\lambda = (2\pi^2 k_{B} m^*)/(\hbar eB)$. $\hbar$ and $k_\text{B}$ are the reduced Plank's constant and the Boltzmann constant, respectively. $T_\text{D}$ is the Dingle temperature, $f$ is the QO frequency, $2\pi \beta$ is the Berry phase and $\delta$ is a phase shift factor. The quantum life time ($\tau_q$) and mobility ($\mu_q$) were calculated by $\tau_q=\hbar/2\pi k_\text{B}T_\text{D}$ and $\mu_q=e\tau_q/m^*$.

\noindent\textbf{Density-functional calculations.} 
DFT calculations were performed by using the code of Vienna ab-initio simulation package (VASP) \cite{kresse1996efficient}, with the experimental lattice parameters and atomic positions (\textcolor{black}{Supplementary Table S2}) as the input. To account for the localized f-electrons, an on-site Hubbard  $\text{U}=5\,$eV  was applied on Eu-4f orbitals \cite{dudarev1998electron}. The calculated magnetic moment is around 6.9 $\mu_\text{B}$/Eu, close to the experimentally measured one. The DFT electronic band structure and magnetization were double-checked by the full-potential local-orbital code (FPLO) with localized atomic basis and full potential \cite{koepernik1999full}. To calculate the Fermi surfaces, we projected Bloch wavefuctions onto maximally localized Wannier functions (MLWFs) \cite{mostofi2008wannier90} and tight-binding model Hamiltonians were constructed from the MLWFs overlap matrix. By performing constant energy contour slices, we were able to obtain the extremal cross-sectional area which is related to the frequency of each pocket as a function of angle and can be used to compare with the quantum oscillation measurements.


\section{acknowledgments}

We are grateful to J. Cano, N.P. Ong, P. Dai and D. Natelson for fruitful discussions.
KA, SL and EM acknowledge support from AFOSR Grant no. FA9550-21-1-0343. JMM has been supported by the NSF Graduate Research Fellowship Program under Grant no. DGE-1842494 and National Science Foundation (NSF) DMR Grant no. 1903741. A portion of this work was performed at the National High Magnetic Field Laboratory, which is supported by NSF Cooperative Agreement no. DMR-1644779 and the State of Florida. This research used resources of the Advanced Light Source and the Stanford Synchrotron Radiation Lightsource, both U.S. Department of Energy (DOE) Office of Science User Facilities under contract nos. AC02-76SF00515 and DE-AC02-05CH11231, respectively. \textcolor{black}{The work at LBL and UC Berkeley was funded by the U.S. DOE, Office of Science, Office of Basic Energy Sciences, Materials Sciences and Engineering Division under contract no. DE-AC02-05-CH11231 (Quantum Materials program KC2202)}. ARPES work at Rice is supported by the Gordon and Betty Moore Foundation’s EPiQS Initiative through grant no. GBMF9470 (JH and MY) and the AFOSR Grant no. FA9550-21-1-0343 (YZ). \textcolor{black}{TCW and MSF acknowledge support from Welch Foundation Grant no. C-1809}. LB is supported by the US-DOE, BES program, through award no. DE-SC0002613 for measurements performed under high magnetic fields.

\section{Author Contributions}
S.L. and E.M. conceived the project. J.M.M. grew the crystals. S.L. and K.A. performed the transport and SdH measurement using a lab magnetometer. B.C. and L.B. performed the high-field transport measurement. \textcolor{black}{T.C.W. and M.S.F. developed the magnetotransport model}. S.L. performed the SdH analysis and K.A. did the XRD measurement and performed the Rietveld structure refinement. Y.S. calculated the DFT electronic structures and performed the symmetry analysis. S.L., Y.S. and K.A. did the extremal cross sectional analysis of the theoretical Fermi surface. ARPES experiments were carried out by JH, YZ, \textcolor{black}{JSO}, MY and \textcolor{black}{RB} with assistance from MH, DL, JD, \textcolor{black}{CJ, AB, and ER}. All authors discussed the results and contributed to writing the manuscript.

\section{Competing Interests}The authors declare no competing interests.

\section{Data availability} 
The data that support the findings of this study are available from the corresponding author upon reasonable request.


\end{document}


\maketitle
\\
${}^*$These authors contributed equally to this work.\\
\noindent ${}^\#$Correspondence to: sl160@rice.edu, sunyan@imr.ac.cn, emorosan@rice.edu
\\
\\
\textbf{Contents}
\vspace{-5mm}
\begin{enumerate}
\begin{spacing}{1.0}
\item Magnetic phase diagram of EuGa$_4$
\item Spinless nodal lines in the paramagnetic EuGa$_4$
\item \textcolor{black}{Weyl NR state and Dirac/Weyl point states in square-net materials}
\item Photon energy dependent data \textcolor{black}{and more ARPES spectra}
\item DFT predicted Fermi surface of EuGa$_4$ in the SP phase
\item Quantum oscillation measurements
\item $\gamma$ pockets in the SP phase and their extremal cross-sectional orbits
\item Cyclotron effective mass of the $\gamma$ pockets
\item Large, non-saturating MR in EuGa$_4$
\item MR value of EuGa$_4$ compared to that of other known topological semimetals
\item \textcolor{black}{Carrier density of EuGa$_4$}
\item \textcolor{black}{Magnetotransport theory of Weyl nodal-ring semimetals}
\item Structural refinement from powder x-ray diffraction

\end{spacing}
\end{enumerate}

\newpage
\begin{spacing}{1.2}
\section{\label{sec1}Magnetic phase diagram of EuGa$_4$} 
\begin{figure*}[htb]
  \includegraphics[width=1\textwidth]{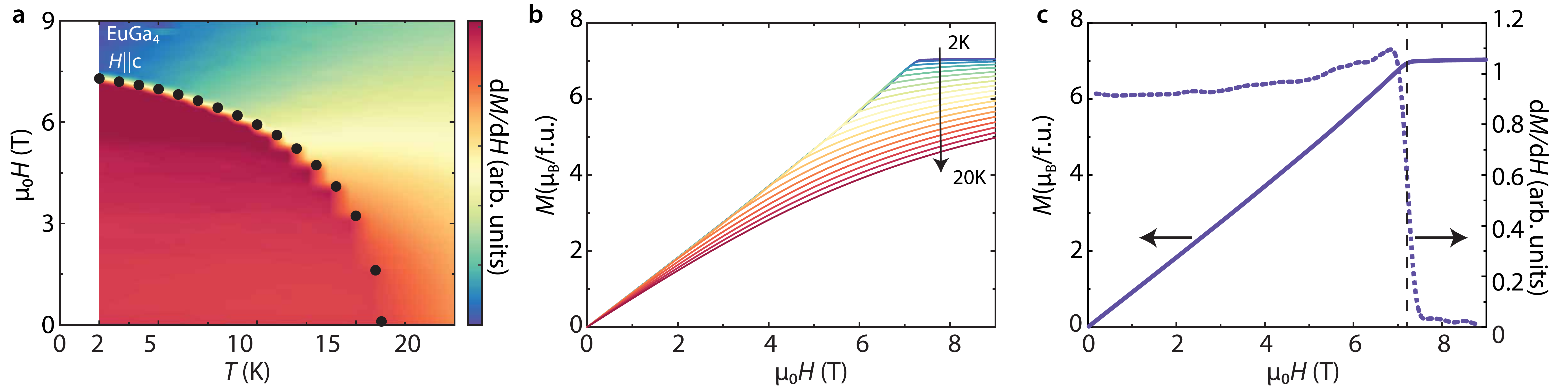}
\caption{\textbf{a}, $H-T$ magnetic phase diagram for EuGa$_{4}$ with magnetic field $H\parallel c$, where the contour plot represents d$M/\text{d}H$ values. \textbf{b}, \textit{M}(\textit{H}) curves measured with temperatures from $2\,$K to $20\,$K up to $\mu_{0}H = 9\,$T. \textbf{c}, Field dependent magnetization at $2\,$K  along with the d$M/\text{d}H$ curve where the dashed vertical line indicates the critical field for magnetic phase transition.}
  \label{figureS1}
\end{figure*}

Figure~\ref{figureS1}a shows the $H-T$ phase diagram determined from isothermal magnetization measurements. The measured $M$($H$) curves at different temperatures are presented in Fig. \ref{figureS1}b. The magnetic phase boundary (dots in Fig.~\ref{figureS1}a) is determined from the derivative  d$M/\text{d}H$, as exemplified in Fig.~\ref{figureS1}c (dashed line, right axis). Note that the magnetic moment saturates to $7~\mu_B/$Eu in the spin-polarized (SP) phase above $\mu_\text{0}H=7.4\,$T at $2\,$K.

\section{\label{sec2}Spinless nodal lines in the paramagnetic EuGa$_4$}

\begin{figure*}[htb]
\centering
  \includegraphics[width=0.8\textwidth]{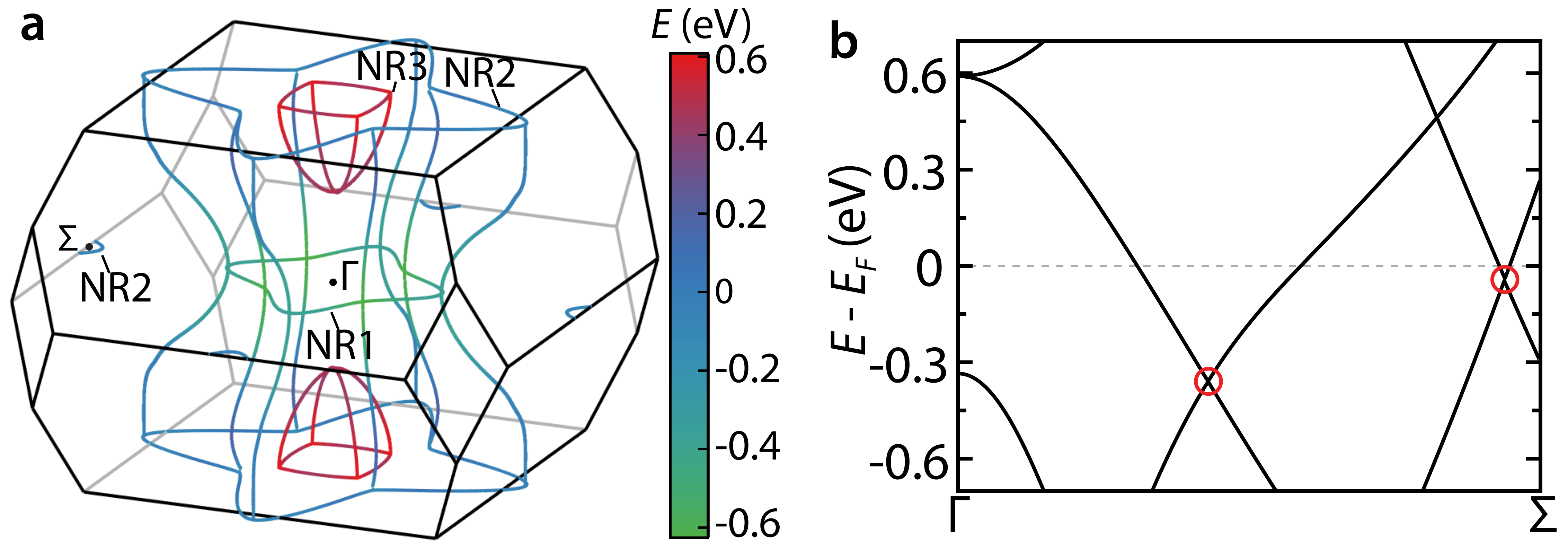}

\caption{The spinless NL state in paramagnetic state of EuGa$_4$. \textbf{a}, The NL network formed by the spinless NLs based on DFT calculations. SOC is not included. Color maps the energy of the NLs. Legend is shown on the right. The NLs in the $kz=0$ and $kz=\pm 2\pi/c$ planes all form a closed NR geometry, with small energy dispersion along the ring. NR1, NR2, and NR3 denote three different NRs. \textbf{b}, The band structure along the $\Gamma-\Sigma$ path. The two nodes below $E_F$ are circled. They extend to form the NR1 and NR2 in the 3D \textit{k}-space.}
  \label{figureS2}
\end{figure*}

Figure~\ref{figureS2}a shows the spinless NLs in the paramagnetic (PM) state for EuGa$_4$, in the absence of SOC. The NLs in the $k_z=$ 0 and $k_z=\pm 2\pi/c$ planes all form closed NRs (indicated by NR1, NR2, and NR3). Along the $\Gamma-\Sigma$ path, there are two crossings below $E_F$ (circled in Fig.~\ref{figureS2}b), as discussed in the main text. These two nodes extend to form two NRs (NR1 and NR2) in the 3D $k$ space. Note that only small parts of the NR2 on the $k_z=\pm 2\pi/c$ planes reside outside of the BZ.
However, for this specific type of BZ which is associated with the body centered tetragonal cell, symmetry dictates that the electronic structure on the $k_z=\pm 2\pi/c$ planes in the neighboring BZ is the same as that on the $k_z=0$ plane in the original BZ. Therefore, the majority part of the NR2 lives on the $k_z=\pm 2\pi/c$ planes of the original BZ, while the other parts are folded onto the $k_z=0$ plane by symmetry. ARPES spectra along the $\Gamma-\Sigma$ path are able to access the information on the nodes from both NR1 and NR2. 

\vspace{-12pt}
Compared with the NR1 and NR2, the energy of NR3 is above $E_F$. Interestingly, the NR3 connects two additional NRs that extend in the $k_x=\pm k_y$ planes, forming a cage-shaped network. The \textit{topological nodal chains} with two NLs touching were first proposed in non-symmorphic crystals [\textcolor{blue}{1}], but were also investigated later in a theoretical work on symmorphic crystals [\textcolor{blue}{2}]. The bands that form the NR1, NR2 and NR3 lead to the formation of three groups of Fermi surface (FS) pockets, as shown in Fig.~\ref{figureS2}a. These NR bands in the PM state undergo band splittings in the spin-polarized (SP) state, resulting in the formation of the three pairs of Weyl NRs. 

\section{\label{sec3}\textcolor{black}{Weyl NR and Dirac/Weyl point states in square-net materials}}

\textcolor{black}{Square-net compounds with conduction bands derived from $p_x/p_y$ orbitals 
are known to be a material platform to host the Dirac nodal lines (four-fold degenerate considering the spin degree of freedom) in the absence of SOC [\textcolor{blue}{3}, \textcolor{blue}{4}]. Among the square-net topological semimetals, materials with the formula of MXZ and MXZ$_2$ are most heavily studied, as discussed in the recent review article [\textcolor{blue}{4}]. To date, these studies have been mainly on the Dirac and Weyl point states and their associated physical properties. SrMnBi$_2$ [\textcolor{blue}{5}, \textcolor{blue}{6}] and YbMnBi$_2$ [\textcolor{blue}{7}] are two exemplary MXZ$_2$ compounds that were studied. The former one (antiferromagnetic ground state [\textcolor{blue}{8}] features anisotropic Dirac band dispersions. However, SOC opens a small gap of $\sim40$ meV at the Dirac point [\textcolor{blue}{5}]. For the latter, the spin-degeneracy can be lifted by the spin canting in the canted antiferromagnetic phase, and the band structure calculation points to the realization of Weyl point state [\textcolor{blue}{7}]. Compared to the Dirac/Weyl point states studied in these earlier works, here we aim to establish a different topological semimetal state, Weyl nodal line, where the spinful conduction and valence bands cross along curves in momentum space, rather than at discrete points [\textcolor{blue}{9}, \textcolor{blue}{10}]. In addition, the Weyl NR state in EuGa$_4$ is robust against SOC.}

\section{\label{sec4}Photon energy dependent ARPES data \textcolor{black}{and more ARPES spectra}}

\begin{figure*}[hbp]
\centering
  \includegraphics[width=1\textwidth]{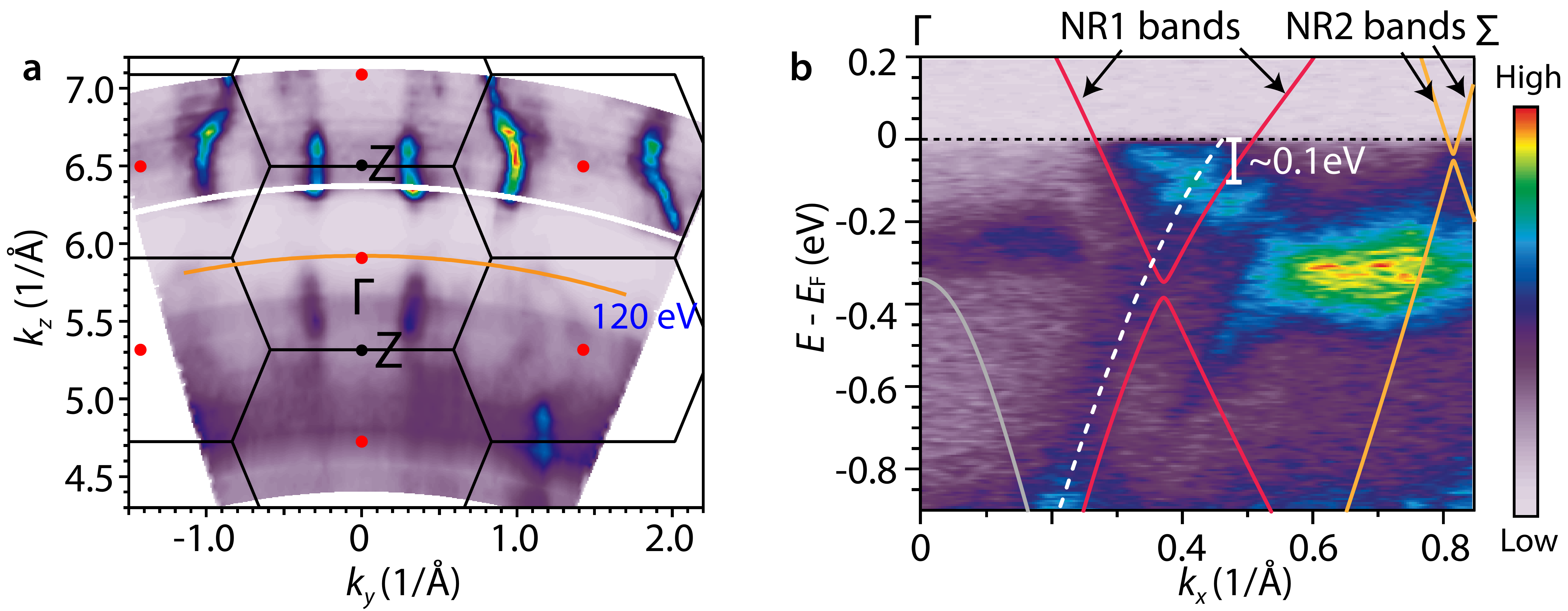}
\caption{\textbf{a}, $k_z$ dependent ARPES intensity along the $\overline{\Sigma}$-$\overline{\Gamma}$-$\overline{\Sigma}$ path, measured with varying photon energy. The band dispersion along the $\Sigma-\Gamma-\Sigma\left(\text{S}\right)-\text{Z}$ path shown in Fig. 2c in the text is measured with photon energy of $h\nu=120\,$eV. \textbf{b}, Band image along $\Gamma-\Sigma$.
The dashed white line delineates a measured band, which is about $0.1\,$eV higher than the corresponding branch of the NR1 bands from DFT calculations (red lines). Yellow lines indicate the NR2 bands from DFT calculations.} 
  \label{figureS3}
\end{figure*}

The FS cross section in the $k_y-k_z$ plane is measured by varying the photon energies from 60 to $180\,$eV, as shown in Fig.~\ref{figureS3}a. The band dispersion measured with the photon energy of $120\,$eV corresponds to the $k_z=0$ plane.

\begin{figure*}[htb]
\centering
  \includegraphics[width=0.75\textwidth]{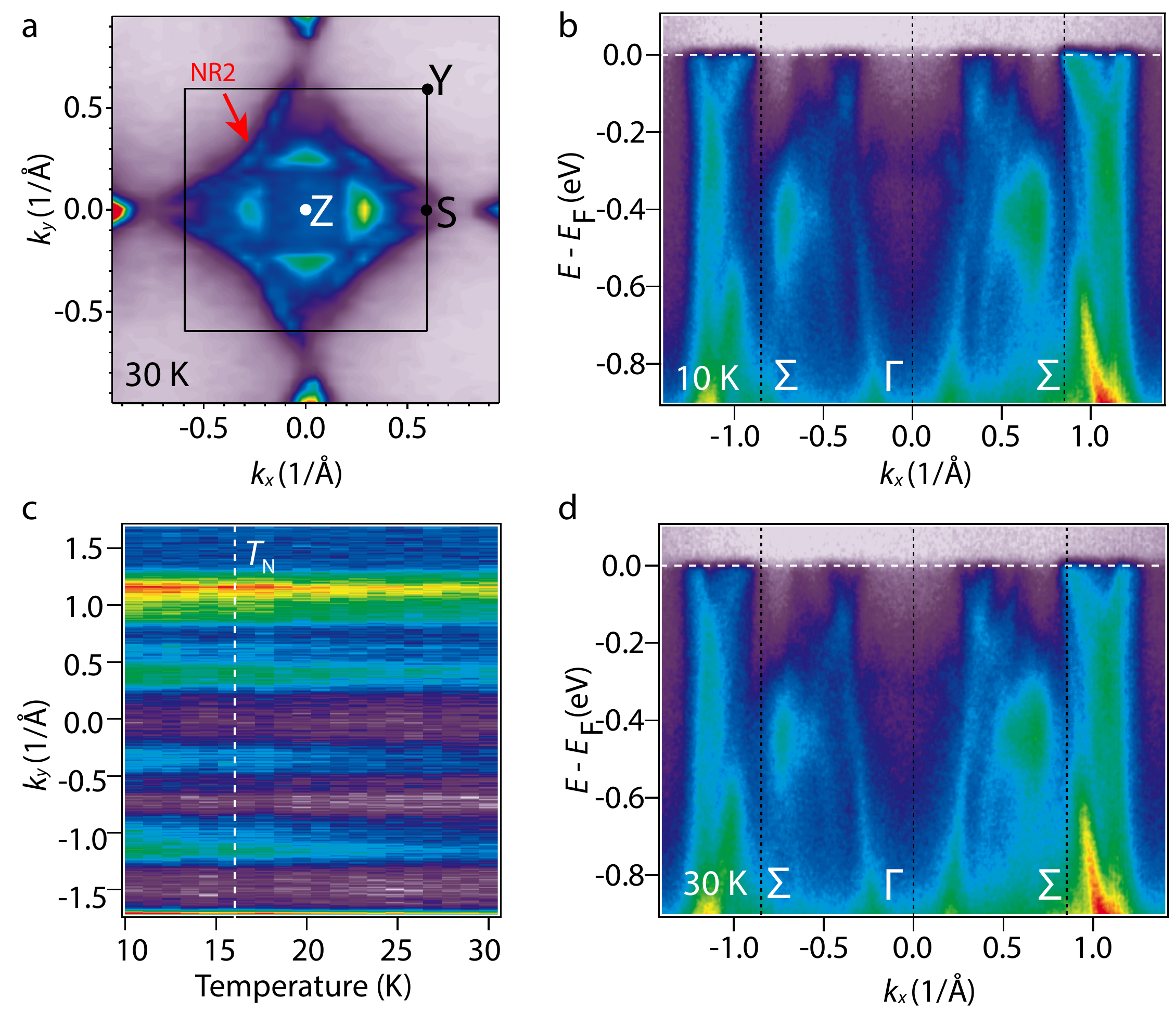}
\caption{\textbf{a}, ARPES measured FS cross section at the $k_z=2\pi/c$ plane with
$h\nu = 146$ eV and $T = 30$ K on a different crystal, featuring the spinless NR2 crossing the Brillouin zone, which is consistent with the DFT prediction (Fig. \ref{figureS2}a). \textbf{b,d,} ARPES band dispersion along the $\Sigma-\Gamma-\Sigma$ path measured at 10~K and 30~K, respectively. \textbf{c,} Constant energy contour of the MDC curves along the $\Sigma-\Gamma-\Sigma$ path at $E_F$, measured at a series of different temperatures ranging from 10~K to 30~K. The dashed line marks the N\'eel temperature, $T_N$.} 
  \label{figure_ARPES}
\end{figure*}

\textcolor{black}{To confirm the reproducibility of the ARPES spectra, we repeated the ARPES measurements on a different EuGa$_4$ single crystal. The measured ARPES FS at the $k_z=2\pi/c$ plane is shown in Fig. ~\ref{figure_ARPES}a, featuring the spinless Dirac NR2, consistent with the DFT prediction (see the illustration in Fig.~\ref{figureS2}a). We also checked the temperature dependent ARPES spectra to evaluate any observable change in the electronic structure from PM to AFM phase. Figures ~\ref{figure_ARPES}b,d show the ARPES spectra measured at 10 K (AFM phase) and 30 K (PM phase). Unfortunately, we are not able to identify any change in the electronic structure. We also measured the temperature dependent momentum distribution curves (MDCs) and plot its constant energy contour at $E_F$ in Fig. ~\ref{figure_ARPES}c; no clear changes can be identified across the AFM transition either. Future high-resolution ARPES experiments, such as laser-ARPES, focusing on the electronic structure around $E_F$ would be crucial in revealing the subtle electronic structure change resulting from the AFM order.}

\section{\label{sec5}DFT predicted Fermi surface of EuGa$_4$ in the SP phase}

\begin{figure*}[htb]
\centering
  \includegraphics[width=0.9\textwidth]{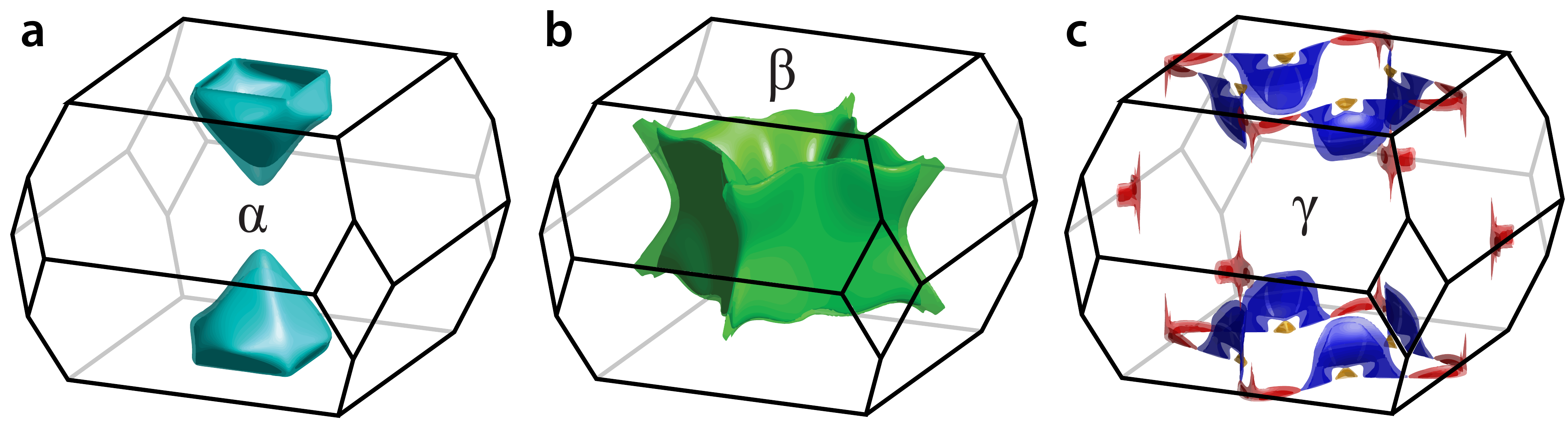}
\caption{Three groups of Fermi surface pockets of \mbox{EuGa$_4$} in the SP phase.}
  \label{figureS5}
\end{figure*}

The bands that form the NR1, NR2, and NR3 (Fig.~\ref{figureS2}a) in the PM phase of EuGa$_4$ undergo band splittings in the SP phase, which lead to the formation of three groups of FS pockets, as shown in Fig.~\ref{figureS5}. Each group of FS pockets appear in pairs. 

\section{\label{sec6}Quantum oscillation measurements}

\begin{figure*}[htb]
  \includegraphics[width=1\textwidth]{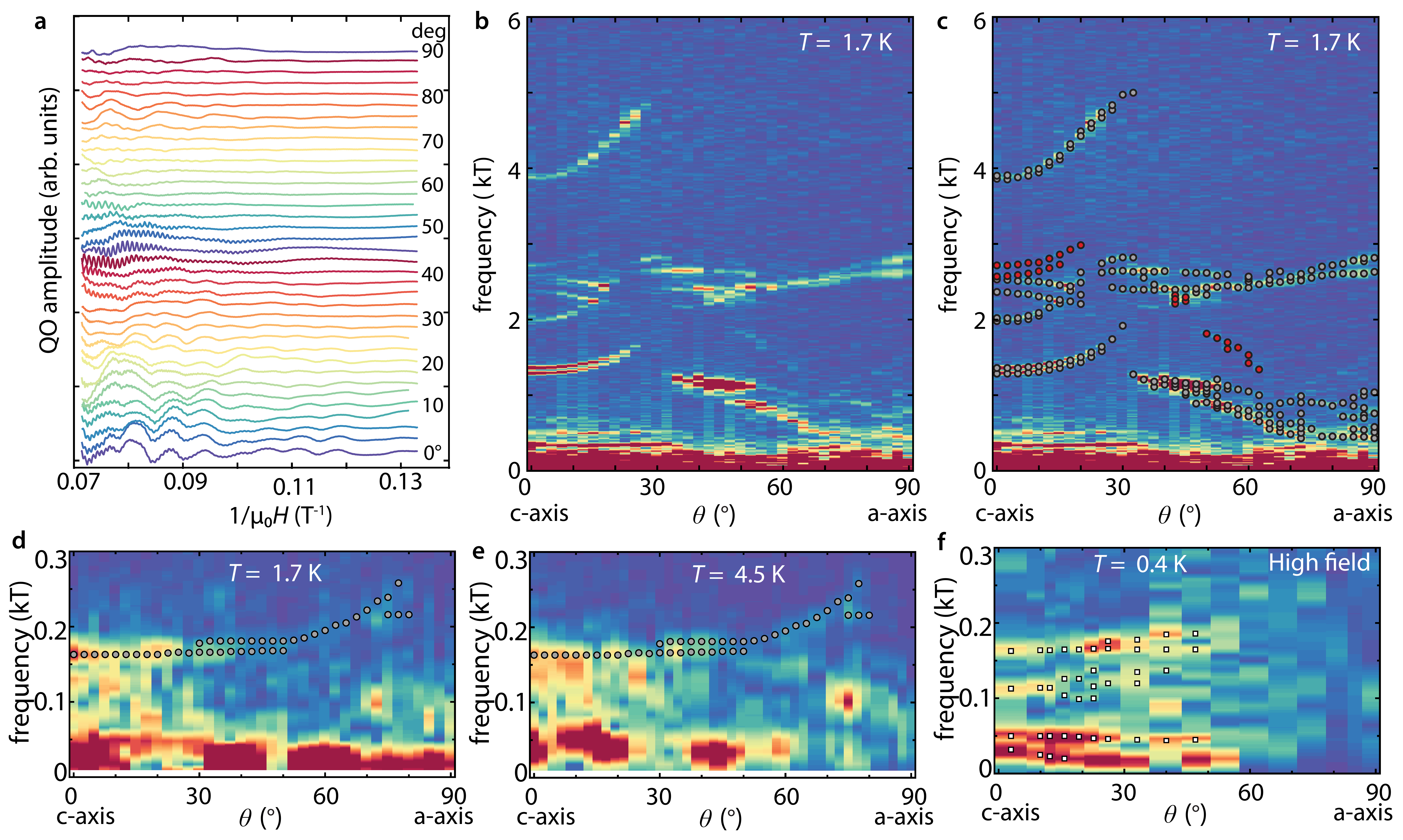}
\caption{Angle dependent SdH oscillations at $T=1.7\,$K, in the SP phase of EuGa$_4$. \textbf{a,} Quantum oscillation after background subtraction at various angles from $0^\circ$ to $90^\circ$. The curves are purposely vertically shifted for better visualization. \textbf{b,c,} Contour plot of the FFT intensity of the QOs at each angle, and the same plot with peak locations marked by the circles. Note that the red circles indicate the harmonic frequency. \textbf{d-f,} Contour plots of the FFT intensity of the QOs at low frequencies. Note that panel (\textbf{d}) is the zoom-in view of the low-frequency region of panel (\textbf{b}). Panel (\textbf{e}) is measured at $T=4.5\,$K. Panel (\textbf{f}) is the one measured with fields up to $41.5\,$T at $T=0.4\,$K.}
  \label{figureS6}
\end{figure*}

SdH oscillations measured in our lab magnetometer (up to $14\,$T) at various angles are sampled with a small angle increment of $2.5^\circ$ from $0^\circ$ to $90^\circ$, as shown in Fig.~\ref{figureS6}a. The contour plot of the fast Fourier transform (FFT) intensity is shown in Fig.~\ref{figureS6}b. The peak locations are then extracted and marked as circles, as shown in Fig.~\ref{figureS6}c-e. In addition to the measurements in a lab magnetometer, we have also performed SdH measurements on a separate sample in high field facilities up to $41.5\,$T, but with bigger angle increment. The contour plot of the FFT intensity is shown in Fig.~\ref{figureS6}f. Note that the measurement geometries are slightly different between the high field and lab magnetometer measurements. For the former, the current is applied along the $a$-axis ($j\parallel a$), while the field is rotating in the $a-c$ plane of the sample. For the latter, the current is applied along the $b$-axis ($j\parallel b$), which is always perpendicular to the field rotation plane ($a-c$ plane of the sample). In the high-field measurement, the MR response is significantly reduced when the field is rotated to approach the current direction. The data become noisy when the rotation angle is larger than $55^\circ$, making it difficult to extract the FFT peaks. Nevertheless, two clear trends of QO frequency evolution can be identified up to $\sim50^\circ$ and two other trends up to $\sim15^\circ$, as marked by the square symbols in Fig.~\ref{figureS6}f. QO frequencies with low FFT amplitude and broadened peaks are not marked due to a difficulty in identifying the peak locations. QO frequencies below $\sim20\,$T are not marked either due to the limited resolution of measurements.

\vspace{-12pt}
With the lab magnetometer measurements, we identified the angular evolution of QO frequencies for the $\alpha$ and $\beta$ pockets. Combining the results from the lab magnetometer and the high-field measurements, we identified the evolution of the $\gamma_4$ frequency, which is consistent with the predicted features based on the outer blue $\gamma$ pocket shown in Fig.~\ref{figureS5}c. In addition to the $\gamma_4$ frequency, both lab magnetometer and the high-field SdH measurements reveal multiple smaller frequencies. 
Based on the DFT calculations, there are indeed multiple extremal cyclotron orbits for the $\gamma$ pockets (see Fig.~\ref{figureS7}), as we will discuss in the next section. The resulting QOs frequencies are packed in a small window. Additionally, there are harmonics and magnetic breakdown with these low-frequency QOs. These factors make the correct identification of the origin of these lower-frequency QOs more difficult.  A further investigation up to higher fields and with a finer step size will help.

\vspace{-12pt}
We notice that de Haas-van Alphen oscillation measurements were performed to probe the Fermi surface geometry in EuGa$_4$ in a prior study [\textcolor{blue}{11}]. Unfortunately, the QOs with frequencies $f>2500\,$T were not resolved when $H\parallel c$. Therefore, this work was not able to identify the $\alpha_\text{belly}$ and $\beta_\text{out}$ frequencies. Furthermore, the small $\gamma$ pockets and their QO features were not resolved or discussed. The topological characters of the bands that lead to the formation of these pockets were unknown. 

\section{\label{sec7}$\gamma$ pockets in the SP phase and their extremal cross-sectional orbits}

Based on the ARPES measurements on the PM EuGa$_4$, we have concluded that one branch of the bands that lead to the formation of the $\beta$ pockets has slightly higher energy ($\sim0.1\,$eV) than the theoretical prediction (see Fig.~\ref{figureS3}b). Consequently, the outer cross section of the $\beta$ pockets should have smaller area than the theoretical value. This conclusion is consistent with our QO measurements in the SP phase. The measured $\beta_\text{out}$ QO frequencies are $400-600$ T below the theoretical predictions. Assuming a rigid band correction, an upshift of the theoretical bands by $\sim90-100$ meV (illustrated in Fig. 3f) is required to reproduce the measured $\beta_\text{out}$ QO frequencies.
 
\begin{figure*}[hp]
  \includegraphics[width=1\textwidth]{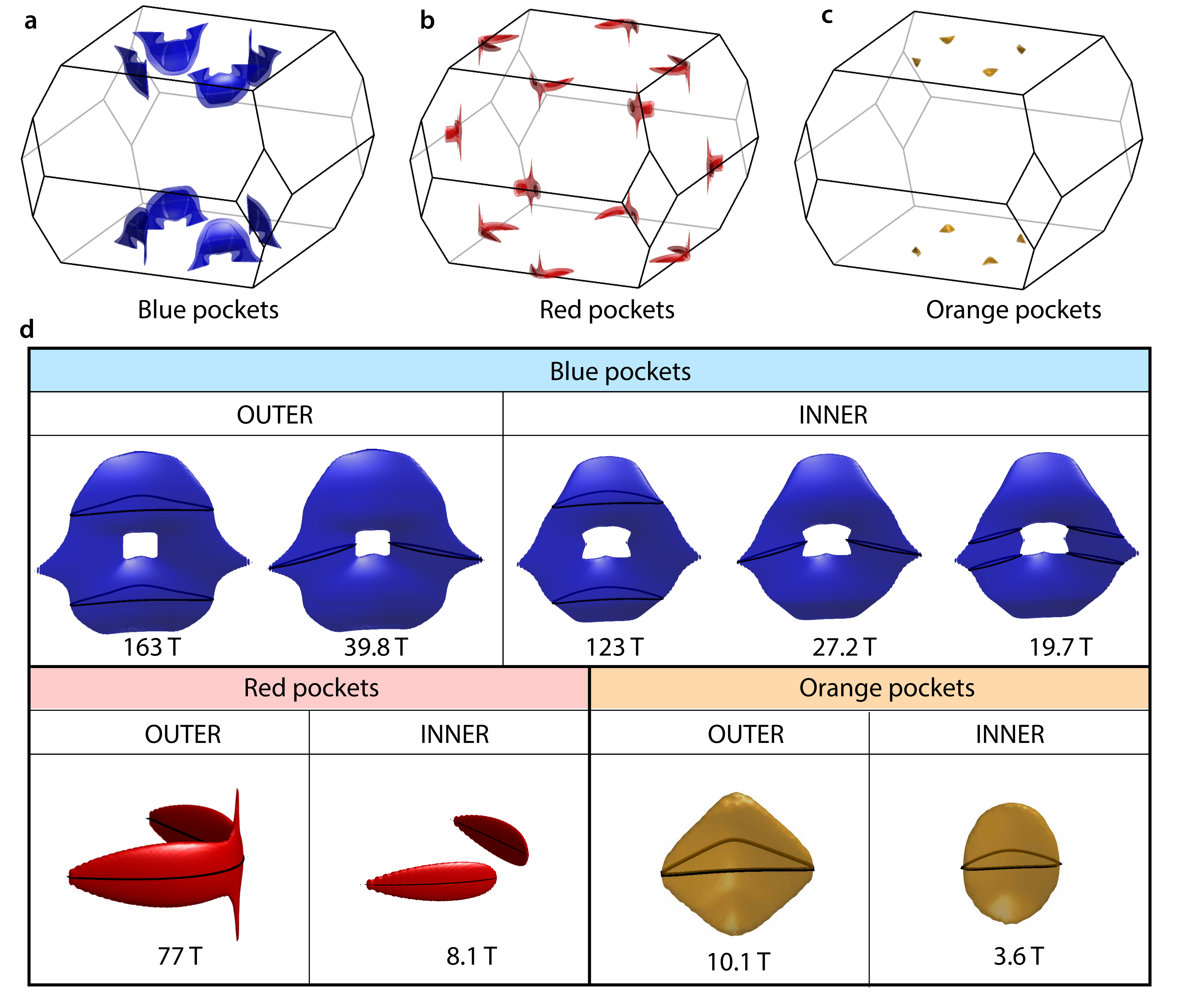}
\caption{Fermi surface pockets of the $\gamma$ pockets. \textbf{a-c,} The blue, red, and orange pockets in the EuGa$_4$ Brillouin zone. Note that the outer pockets are rendered semi-transparent so that the inner ones are revealed. \textbf{d,} Enlarged view of each individual blue, red, and orange pocket. The black lines illustrate the extremal cross-sectional orbits of each pocket when $H\parallel c$. The number below each pocket indicates the QO frequency of the orbit in the unit of Tesla.
}
  \label{figureS7}
\end{figure*}

\begin{figure*}[h]
\center
  \includegraphics[width=0.8\textwidth]{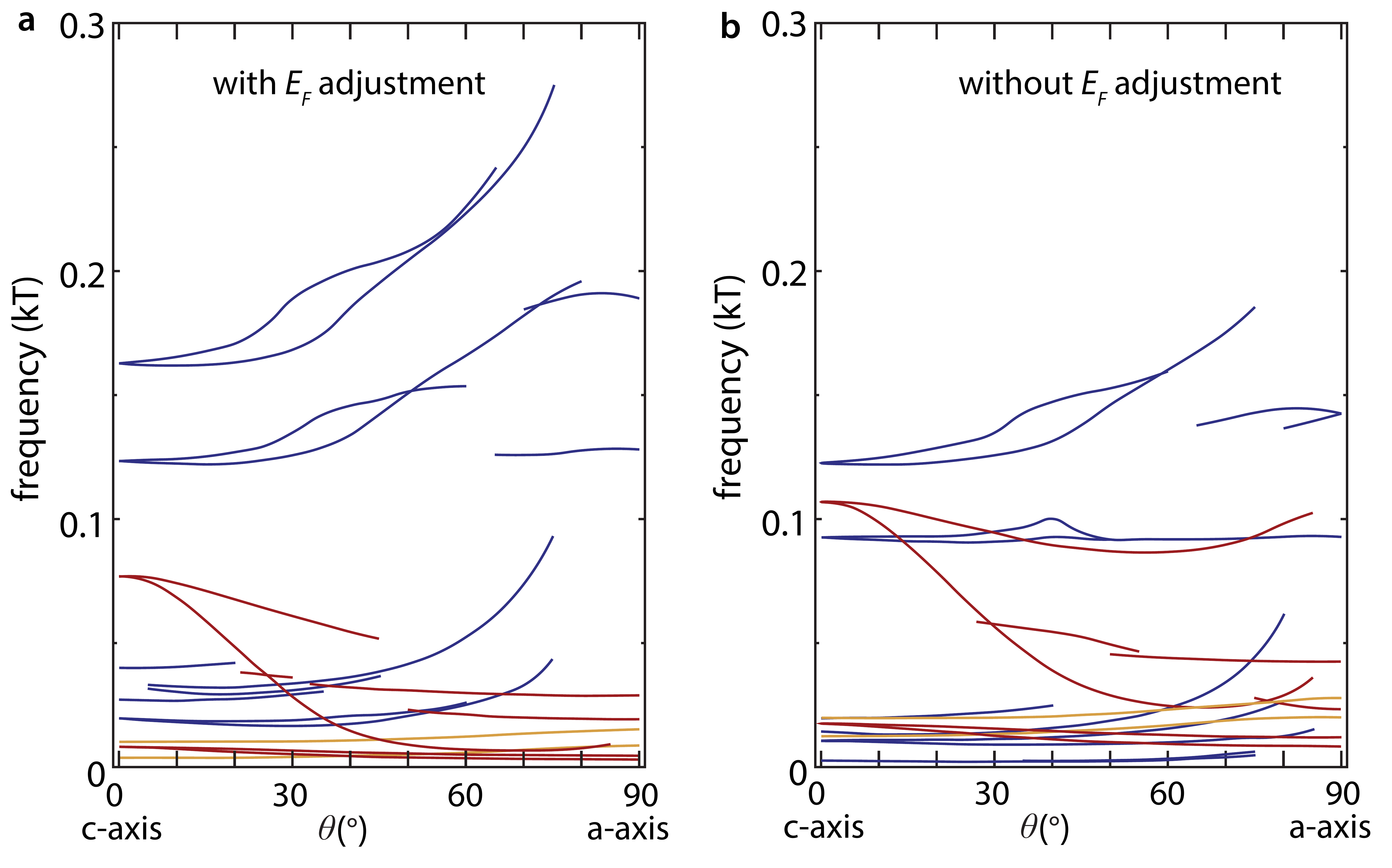}
\caption{\textbf{a,b,} Theoretically predicted angle dependent oscillation frequency of $\gamma$ pockets, in the SP phase ($m\parallel c$) of EuGa$_4$, with and without $E_F$ adjustment, respectively.}
  \label{figureS8}
\end{figure*}
 

The knowledge of the upshift energy correction to the $\beta_\text{out}$ bands is helpful for the determination of the energy correction to the DFT bands that form the $\gamma$ pockets, since they partially share the same band characters. In particular, the blue hole (red electron) pockets (see the illustration in Fig.~\ref{figureS5}c) will be larger (smaller) than the theoretical predictions, along with the upshift of the $\beta_\text{out}$  bands. However, the exact value of the energy shift of the bands that form the $\gamma$ pockets can be smaller or larger than $90-100$ meV, due to the existence of $k_z$ dispersion. 
One simple treatment is to rigidly adjust the Fermi energy ($E_F$) of all the pertinent bands that are responsible for the formation of the probed pockets. In reality, however, it is likely that the pockets are formed by two or more bands, but only one of them needs an adjustment while the others do not [\textcolor{blue}{12}]. In this scenario, an $E_F$ adjustment to all the bands by one common number can be considered as an averaging treatment. 

We have evaluated the $E_F$ adjustment to the bands that form the $\gamma$ pockets. In particular, we find that an upshift of the bands, or equivalently a lowering of $E_F$ by $35\,$meV is necessary to reproduce the measured QO frequency of $163\,$T ($\theta=0^\circ$) from the blue pockets (Fig.~\ref{figureS7}). The required upshift of the bands is consistent with the expectation based on the ARPES results. As for the red pockets (Fig.~\ref{figureS7}), our ARPES measurements suggest that the band crossing near the $\Sigma$ point in the BZ is $20\pm10\,$meV higher than the theoretical prediction. Therefore, the $E_F$ of the bands is lowered by $20\,$meV to obtain the theoretical QO frequencies associated with the red pockets (Fig.~\ref{figureS7}).  Overall, our QO measurements indicate that a small upshift of the DFT bands by $20-35\,$meV is necessary to understand the $\gamma$ pockets. Since the $\gamma$ pockets arise from the bands that form the red/blue Weyl NRs (see Fig. 1f,i), we conclude that the energy window of the red/blue NR states are $165-195\,$meV, which is quite small considering it spans the whole $k_z=\pm 2\pi/c$ plane of the Brillouin zone.

\vspace{-12pt}
In Fig.~\ref{figureS7}d, we show all the possible extremal cross-sectional orbits associated with the $\gamma$ pockets in the SP phase of EuGa$_4$ when the field is parallel to the c-axis, based on the DFT calculations. The blue ones are the hole pockets, while the red and oranges ones are the electron pockets. The size of the extremal orbits after $E_F$ adjustment for the blue pockets are: $19.7\,$T, $27.2\,$T, $39.8\,$T, $123\,$T, and $163\,$T. 
Those for the red pockets are $8.1\,$T and  $77\,$T. Those for the orange pockets are $3.6\,$T and $10.1\,$T. Their angle dependent QO frequencies are plotted in Fig.~\ref{figureS8}a. For reference, we also show the angle dependent QO frequencies without any $E_F$ adjustment in Fig.~\ref{figureS8}b.

\section{\label{sec8}Cyclotron effective mass of the $\gamma$ pockets}

\begin{figure*}[h]
   \center
  \includegraphics[width=0.45\textwidth]{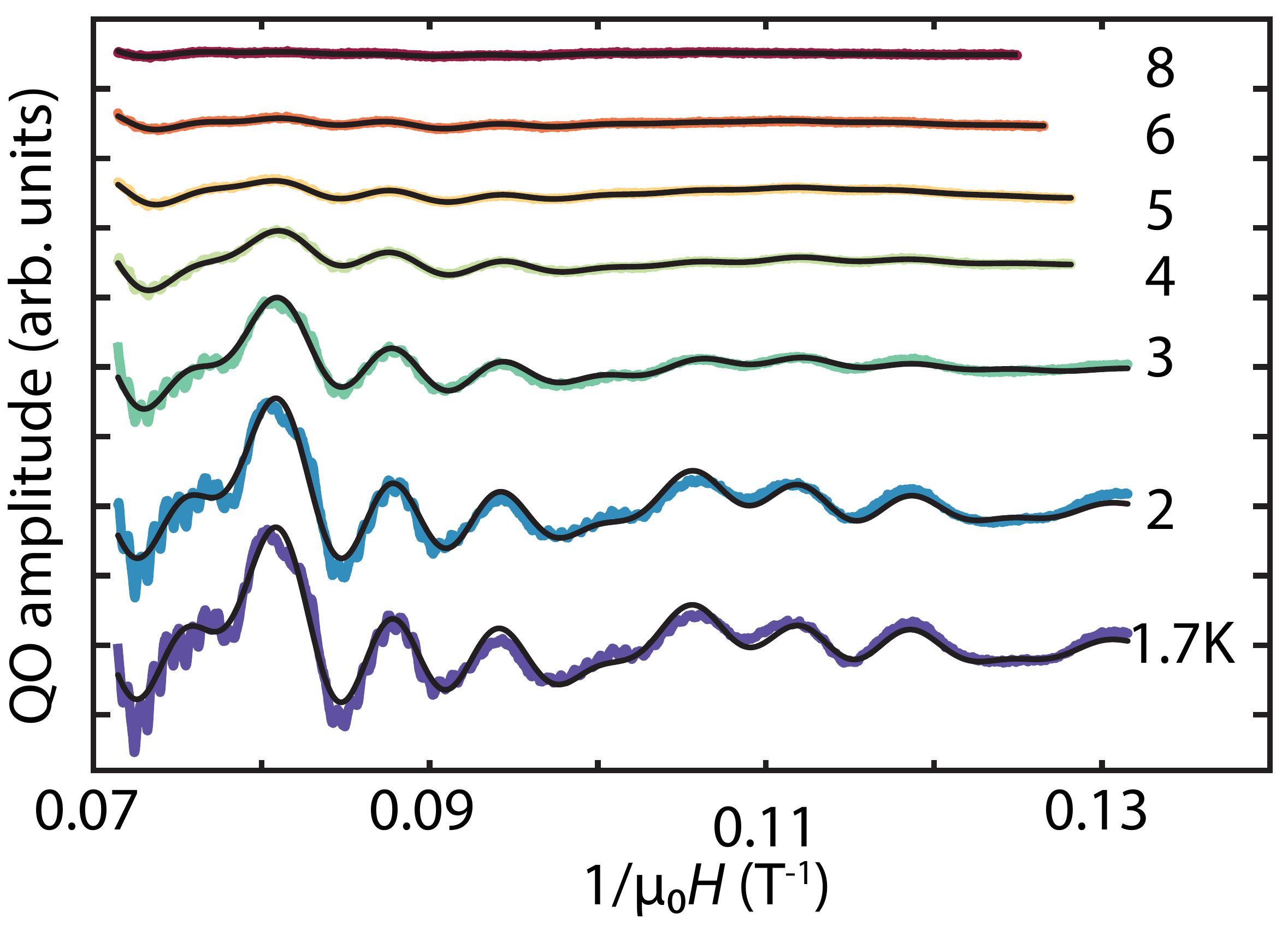}
\caption{Quantum oscillation measured at different temperatures from $1.7\,$K to $8\,$K when $H\parallel c$ ($\theta=0^\circ$). The curves are purposely vertically shifted for better visualization. Note that the solid black lines are L-K fits, based on the four frequency components of $\gamma_1= 30$\,T, $\gamma_2= 77$\,T,  $\gamma_3= 125$\,T, and $\gamma_4= 163$\,T, as discussed in the main text. 
}
\vspace{-12pt}
  \label{figureS9}
\end{figure*}

The measured temperature dependent QOs are presented in Fig. \ref{figureS9}. The L-K fit based on four frequency components are also presented on the top of experimental data. The temperature dependent amplitude of each oscillation component is presented in Fig. 3g in the main text.

Theoretically, the effective mass, $m_\text{eff}$, is expressed as the derivative of the cyclotron orbit area $A$ with respect to the energy $E$ [\textcolor{blue}{13}]:

\vspace{-12pt}
\[m_\text{eff}=\frac{\hbar^2}{2\pi}\frac{\partial A}{\partial E}\].

\vspace{-24pt}
Based on the DFT band structures, $\frac{\partial A}{\partial E}$ can be readily calculated with a small variation of $E_F$, thus providing a way to evaluate the effective mass for each FS pocket in the single-particle frame without correlation effects. As discussed in the main text and Section 6, the QO frequency $\gamma_4$ is associated with the outer blue $\gamma$ pocket (Fig. \ref{figureS5}c), while the nature of the measured lower frequencies is not clearly identified. For analysis, we have calculated the effective masses of all the possible extremal cyclotron orbits that were illustrated in Fig. \ref{figureS7}d. For the two types of the extremal orbits of the outer blue pocket, the effective masses are $0.14m_e$ and $0.075m_e$, where $m_e$ is the mass of an electron. For the three types of the extremal orbits of the inner blue pocket, the effective masses are $0.11m_e$, $0.062m_e$, and $0.064m_e$. For the outer and inner red pockets, the effective masses are $0.18m_e$ and $0.05m_e$, respectively. For the outer and inner orange pockets, the effective masses are $0.03m_e$ and $0.02m_e$, respectively.
Apparently, they are all smaller than those ($0.68-0.76m_e$) determined from experiments. We thus conclude that electronic correlation plays a role for the enhanced effective masses.

\section{\label{sec9} Large, non-saturating MR in EuGa$_4$}

 \begin{figure*}[h]
  \includegraphics[width=1\textwidth]{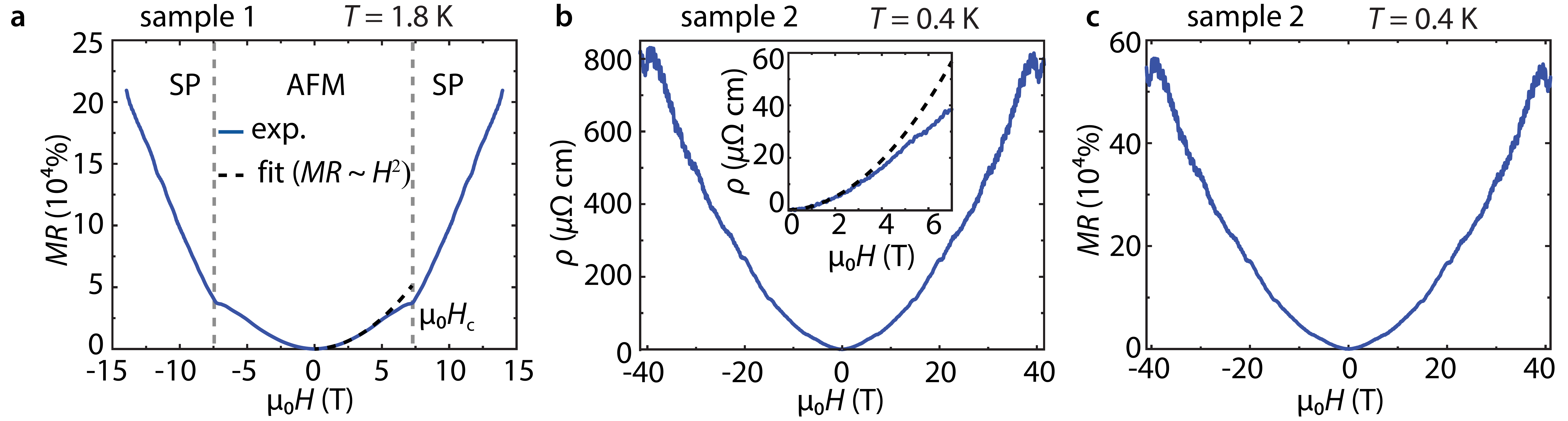}
\caption{MR behaviors of EuGa$_4$. \textbf{a}, MR curves measured in lab magnetometer up to $14\,$T on sample 1. $H^2$ fit is performed from 0 to $3.5\,$T. The AFM-SP phase transition is marked by $\mu_0H_c$. \textbf{b}, Field dependent resistivity from $-41.5\,$T to $41.5\,$T at $0.4\,$K measured on sample 2. Inset shows the $H^2$ fit to the low-field resistivity up to $2.5\,$T.
\textbf{c}, The high-field MR curve converted from (b). 
}
  \label{figureS10}
\end{figure*}

We performed field-dependent in-plane resistivity measurement on EuGa$_4$ both in our lab magnetometer (up to $14\,$T at $1.8\,$K) and using the high-field facility (up to $41.5\,$T at $0.4\,$K). Two samples are involved in the measurements. The measured resistivity in the positive- and negative-field sweeps is shown in Fig.\,\ref{figureS10}a,b. The MR curves on both samples show signatures of AFM-SP transition; the transition field is marked by arrows in Figs. 4b,c.
To avoid possible sample heating issues from contacts at $T=0.4\,$K during the high-field measurement, we intentionally applied a small current $j=3\,$mA, which results in a low signal-noise-ratio at the low-field regime. In particular, the zero-field resistance reading has a large variation. Since the low-field MR behavior can be nicely described by an $H^2$ relation, an $H^2$ fit to the low-field region (inset, Fig.~\ref{figureS10}b) is performed and the zero-field resistivity is obtained by the fit. Accordingly, we obtain the MR curve for the high-field measurement, as shown in Fig.~\ref{figureS10}c. At 0.4\,K and $\sim40\,$T, the MR exceeds $0.5\times10^6\%$. Note that the resistivity dip at $40\,$T is from quantum oscillation. The MR measurements from our lab magnetometer and high-field facility show clear deviation from the $H^2$ relation below $H_c$. The field locations where the deviation occurs are at $\sim3.5\,$T and $\sim2\,$T for the former and latter measurements, respectively. The difference might be due to a slight difference in the sample quality.

\section{\label{sec10}MR value of EuGa$_4$ compared to that of other known topological semimetals}
Table ~\ref{tab:my_label} lists the non-magnetic and magnetic topological semimetals and their MR values that are included for comparison with EuGa$_4$. The plot is shown in Fig. 4e.

\begin{table}[ht]
    \centering
    \caption{List of TSMs with their respective MR, temperature (\textit{T}) and field ($\mu_{0}H$) conditions.}
    \vspace{0mm}
    \begin{tabular}{c|c|c|c|c}
         Compound & ~MR (\%) & ~\textit{T} (K)~ & $~\mu_0H$ (T)~ & Reference \\
         \hline
         Na$_{3}$Bi & 529 & 4.5 & 9 & [\textcolor{blue}{14}]\\
         TaAs & 70,000 & 1.8 & 9 & [\textcolor{blue}{15}]\\
         Cd$_{3}$As$_{2}$ & 133,600 & 5 & 9 & [\textcolor{blue}{16}]\\ 
         WTe$_{2}$ & 452,700 & 4.5 & 14.7 & [\textcolor{blue}{17}]\\ 
         NbP & 850,000 & 1.85 & 9 & [\textcolor{blue}{18}]\\
         Co$_{3}$Sn$_{2}$S$_{2}$ & 53 & 2 & 14 & [\textcolor{blue}{19}]\\
         Fe$_{3}$Sn$_{2}$ & 88.6 & 0.6 & 14 & [\textcolor{blue}{20}]\\
         HgCr$_{2}$Se$_{4}$ & 100 & 2 & 8 & [\textcolor{blue}{21}]\\ 
         MnBi & 250 & 2 & 9 & [\textcolor{blue}{22}]\\
         PrAlSi & 314 & 2 & 9 & [\textcolor{blue}{23}] \\
         CeAlGe & 6 & 2 & 7 & [\textcolor{blue}{24}]\\ 
         NdPtBi & 75 & 2 & 9 & [\textcolor{blue}{25}]\\
         HoPtBi & 122 & 2 & 14 & [\textcolor{blue}{26}]\\
         GdPtBi & 150 & 2 & 9 & [\textcolor{blue}{27}]\\
         FeSn & 224 & 0.4 & 14 & [\textcolor{blue}{28}]  \\
         SrMnBi$_{2}$ & 291 & 2 & 14 & [\textcolor{blue}{5}]\\
         TbPtBi & 392 & 2 & 14 & [\textcolor{blue}{26}]\\
         NdAlSi & 454 & 2 & 14 & [\textcolor{blue}{29}]\\
         YbMnBi$_2$ & 573 & 2 & 9 & [\textcolor{blue}{7}]\\
         EuGa$_{4}$ & 210,000 & 2 & 14 & This work.\\
         \hline
    \end{tabular}
    \label{tab:my_label}
\end{table}

\section{\label{sec11}Carrier density of EuGa$_4$}

\begin{figure*}[htp]
  \includegraphics[width=0.9\textwidth]{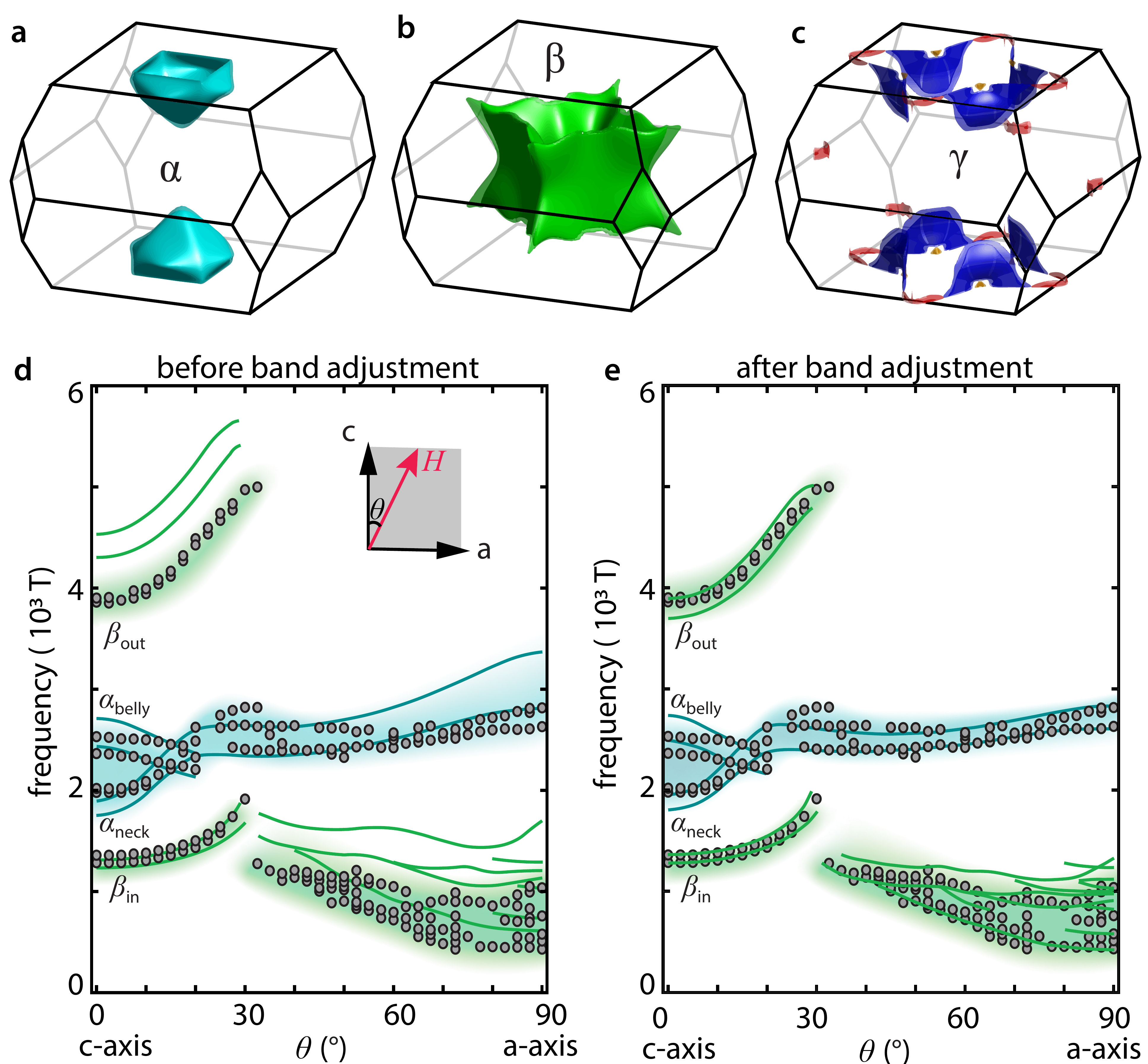}
  \centering
\caption{\textbf{a--c} FS of EuGa$_4$ after band adjustment in the tight-binding model calculations. \textbf{d,e} The angle dependent quantum oscillation data with theoretical prediction before and after the band adjustment. Note that panel (d) is reproduced from Fig. 3c in the main text for the convenience of comparison.
}
  \label{FS_TBM}
\end{figure*}

\textcolor{black}{The DFT predicted Fermi surface reasonably well describes the measured angle dependent quantum oscillation data, except slight overestimates of the outer cross-sections of the spin-split $\beta$-pockets (electron), and the high-angle ($\theta>60^\circ$) cross-sections  of the $\alpha$-pockets (hole) (Fig. 3c). Overall, we expect an overestimate of the carrier density by the DFT calculation. With the experimental quantum oscillation data, we are able to improve the accuracy. To this end, we constructed the tight-binding model Hamiltonian of EuGa$_4$ in the SP phase according to the DFT calculation result. We then selectively adjust the energy of the bands gently to reproduce the experimentally measured quantum oscillation frequencies. Essentially, we projected the Bloch wavefuctions onto maximally localized Wannier functions (MLWFs) [\textcolor{blue}{30}], and the model Hamiltonian was constructed from the MLWFs overlap matrix. In Figs. ~\ref{FS_TBM}a--c, we show the FS plots of the $\alpha$- and $\beta-$pockets after the band adjustment. By comparing these FS pockets with the ones (Figs. ~\ref{figureS5}) without band adjustment, one can see that the FS maintains the same morphology except slight shrinking or distortion. We show the angle dependent quantum oscillation data from experiment and theory, before and after the band adjustment in Figs. ~\ref{FS_TBM}d,e.}

\textcolor{black}{By comparing panel (d) and panel (e) in Fig. ~\ref{FS_TBM}, one can observe that the band adjustment treatment has indeed quantitatively improved the accuracy of the theory calculated Fermi surface. We thus calculate the carrier density associated with each pocket and present the results in Table ~\ref{carrier_density}. The total electron and hole carrier density are determined to be $1.01\times 10^{21}$cm$^{-3}$ and $0.92\times 10^{21}$cm$^{-3}$, respectively. The ratio is thus determined to be $n_e/n_h=1.10$ after the band adjustment. For comparison, $n_e/n_h=1.40$ before any band adjustment. }

\begin{table}[htp]
    \centering
    \caption{Carrier density of EuGa$_4$ in the SP phase.}
    \vspace{0mm}
    
    \setlength{\tabcolsep}{13pt}
\begin{tabular}{|>{\centering\arraybackslash}m{1.5cm}|>{\centering\arraybackslash}m{2cm}|>{\centering\arraybackslash}m{4cm}|>{\centering\arraybackslash}m{4cm}|}
\hline

         Fermi surface & carrier type & carrier density  before band adjustment ($\times10^{20}$cm$^{-3}$) & carrier density after band adjustment ($\times10^{20}$cm$^{-3}$) \\
         \hline
        \multirow{2}{2em}{\hspace{3mm}$\alpha$}  &  \multirow{2}{2em}{\hspace{0mm}hole}  & 3.79 & 3.65\\
        && 4.62& 4.05\\
        \hline
        \multirow{2}{2em}{\hspace{3mm}$\gamma$ (blue)}  & \multirow{2}{2em}{\hspace{0mm}hole}   & 0.33& 0.58\\
        && 0.60& 0.92\\
         \hline
       \multirow{2}{2em}{\hspace{3mm}$\beta$}  & \multirow{2}{2em}{electron}   & 5.87& 4.43\\
        && 7.22& 5.58\\
         \hline
      \multirow{2}{2em}{\hspace{3mm}$\gamma$ (red)}  & \multirow{2}{2em}{electron}   & 0.02& $<$0.01\\
        && 0.13& 0.07\\
         \hline
      \multirow{2}{2em}{\hspace{3mm}$\gamma$ (orange)}  & \multirow{2}{2em}{electron}   & $<$0.01& $<$0.01\\
        && 0.01& $<$0.01\\
         \hline
    \end{tabular}
    \label{carrier_density}
\end{table}

\textcolor{black}{Based on the angle dependent quantum oscillation data (Fig. ~\ref{FS_TBM}e), we can evaluate the errors of $n_e$ and $n_h$. 
For the $\alpha-$FS pockets, the spin splitting is clearly resolved by the measured two branches of quantum oscillation frequencies. In the meanwhile, the spin-splitting also causes a difference in the carrier density associated with the two spin-split pockets. The difference is $0.4\times 10^{20}$cm$^{-3}$ (see Table S2). Consequently, the error of carrier density associated with the $\alpha-$pockets must be smaller than half of the difference, which gives $\Delta n_h \left(\alpha\right) <0.4\times 10^{20}$cm$^{-3}/2=0.2\times 10^{20}$cm$^{-3}$. For the torus-shaped $\beta-$FS, the inner ($\beta_{\rm in}$) and outer ($\beta_{\rm out}$) extremal orbits appear below $\sim30^\circ$. For the $\beta_{\rm in}$ orbits, the experiment and theory match really well. For the $\beta_{\rm out}$ orbits, the lower- and higher-branches of the oscillation frequencies are slightly smaller and higher, respectively, than the experimental ones. On average, $n_h \left(\beta\right)$ estimated from theory should be close to the experimental one. If we take an extremely conservative estimate, the error can be calculated as half of the difference in the carrier density: $\Delta n_h \left(\beta\right) <1.15\times 10^{20}$cm$^{-3}/2=0.58\times 10^{20}$cm$^{-3}$. Compared to the $\alpha-$ and $\beta-$pockets, the $\gamma-$pockets are much smaller in volume. Therefore, $\Delta n_h \left(\gamma\right)$ should be considerably smaller than $\Delta n_h \left(\alpha\right)$ and $\Delta n_h \left(\beta\right)$. Based on the analysis above, we conclude that the dominant source of error in determining $n_e/n_h$ is on the $\beta$-pockets. The error bar is thus determined: $\Delta (n_e/n_h)<6\%$.} 

\textcolor{black}{Overall, the ratio of electron and hole carrier density in the SP phase of EuGa$_4$ is $n_e/n_h = 1.10\pm0.06$. We conclude that the close electron-hole carrier density may play a role for the large MR at low fields, but is not close enough to unity to achieve nonsaturating MR up to $\sim40$~T.
We note that the carrier density evaluated this way is more accurate than Hall measurements based on isotropic two-band model, because of the intrinsic FS anisotropy and multiband nature in EuGa$_4$.}

\section{\label{sec12}Magnetotransport theory of Weyl nodal-ring semimetals}

\textcolor{black}{Consider the following model Hamiltonian for a nodal-ring semimetal:
\begin{eqnarray}
\label{eq:hamiltonian}
	\hh(\textbf{k})
	=
	v_z k_z \hsig_1 
	+
	\xikpara \hsig_3
	-
	\mu
	,
	\qquad
	\xikpara 
	=
	\frac{k_{\parallel}^2}{2m}	- 	E_M
	,
\end{eqnarray}
where 
$k_{\parallel} = \sqrt{k_x^2 + k_y^2}$, $\hsig_i$ is Pauli matrix in the orbital space, $E_M$ is an energy scale controlling the radius of the nodal ring, $m$ is a parameter controlling the in-plane effective mass, and $\mu$ is a parameter controlling the chemical potential.
The eigenenergies are given by
\begin{equation}
	\pm
	\veps_{\textbf{k}}
	=
	\pm 
	\sqrt{
		(v_z k_z)^2 + \xikpara^2
	}
	-
	\mu
	.
\end{equation}
Evidently, when $k_x^2 + k_y^2 = 2m E_M$ and $k_z = 0$, the system is gapless, forming a nodal ring with a radius $k_0 = \sqrt{2m E_M}$. The corresponding band structure is shown in the inset of Fig.~\ref{fig:MR_Kubo_normal_and_lnln}a.}

\textcolor{black}{For this Weyl nodal ring model, the Berry curvature for the conduction band is known to be 
$
	\bm{\Omega}_{+,\textbf{k}} = \pi \,\delta(k_0 - k_{\parallel}) \, \delta(k_z) \, \hat{\phi}
	,
$
which is concentrated along the nodal ring [\textcolor{blue}{31}]. 
The expression for the hole band is similar. 
While such non-trivial topology can give rise to an interesting anomalous transverse current [\textcolor{blue}{31}], it is not our primary concern here. 
This is because the electric field $\textbf{E} \parallel \hat{x}$ and $\textbf{B} \parallel \hat{z}$ in the experiment and our main focus is on computing $\sigma_{xx}$ and $\sigma_{xy}$.}

\subsection{A. Kinetic theory for magnetoconductivity}
\hfill \break
\textcolor{black}{In the following calculation, we assume $\mu > 0$. 
In the presence of space-time uniform external electric and magnetic fields, the kinetic equation governing the distribution function of electrons in the linear response regime is given by [\textcolor{blue}{32}, \textcolor{blue}{33}]
\begin{equation}
	-
	e  E \, v^x_{\textbf{k}} \,  \frac{\partial f_{e,0}}{\partial \veps} 
	-
	eB (v^y_{\textbf{k}} \,  \partial_{k_x} - v^x_{\textbf{k}} \, \partial_{k_y})
	\,
	\delta f_{e}(\textbf{k})
	= 
	I_{\coll}[f_{e}(\textbf{k})]
	,
\end{equation}
where $\textbf{v}_{\textbf{k}} = \nablaK \veps_{\textbf{k}}$ is the velocity of electrons,  $f_{e,0}(\veps) = 1/[1 + \exp(\veps/T)]$ is the Fermi Dirac distribution function, $T$ denotes temperature, and $\delta f_e(\textbf{k})$ describes the deviation from equilibrium. Note that because of the orientation of the external fields ($\textbf{B} \parallel \hat{z}$) in the experiment, the Berry curvature does not enter the kinetic equation [\textcolor{blue}{33}]. 
For quenched onsite impurity potential disorder, the collision integral is given by [\textcolor{blue}{34}]
\begin{equation}
	I_{\coll}[f_{e}]
	=
	2\pi \lamimp
	\int\limits_{\textbf{q}}
	\frac{
		1 + \hat{d}_{\textbf{k}}\cdot \hat{d}_{\textbf{q}}
	}{
		2
	}
	\,
	2\pi \delta(\veps_{\textbf{q}} - \veps_{\textbf{k}})
	\,
	[
		f_{e}(\textbf{q}) - f_{e}(\textbf{k})
	]
	,
\end{equation}
where $\int_{\textbf{q}} = \int \frac{d^3 q}{(2\pi)^3}$, $\textbf{d}_{\textbf{k}} =
(0, v_zk_z, \xikpara)$, $\lamimp$ is a parameter controlling the disorder strength, and the Dirac delta function $\delta(\veps_{\textbf{q}} - \veps_{\textbf{p}})$ imposes energy conservation. The factor $(1 + \hat{d}_{\textbf{p}}\cdot \hat{d}_{\textbf{q}})/2$ arises due to the matrix structure of the Hamiltonian and accounts for the enhancement of forward scattering.  }

\textcolor{black}{The kinetic equation can be solved using the \textit{ansatz}
$
	\delta f_{e}(\textbf{k})
	= 
	eE (\partial_{\veps} f_{e,0})
	\textbf{v}_{\textbf{k}}^{\parallel} \cdot \bm{\kappa}(k)
	,
$
where $\bm{\kappa} = (\kappa_x, \kappa_y)$ is an undetermined function depending only on the norm of $\textbf{k}$, $\phi = \tan^{-1}(k_y/k_x)$, and $k_{\parallel} = \sqrt{k_x^2 + k_y^2}$. 
Performing the $\textbf{q}$ integral in $I_{\coll}[f_{e}]$ and solving the kinetic equation, we find
\begin{equation}
	\begin{bmatrix}
		\kappa_x(k) \\ \kappa_y(k)
	\end{bmatrix}
	=
	\frac{
		\tautr(\textbf{k})
	}{
		1 + [\omega_{c,\sfeff}(\textbf{k}) \tautr(\textbf{k})]^2
	}
	\begin{bmatrix}
		1
		\\
		\omega_{c,\sfeff}(\textbf{k}) \tautr(\textbf{k})
	\end{bmatrix}
	,
\end{equation}
where
$
	\omega_{c,\sfeff}(\textbf{k})
	= 
	\omega_c \, \xikpara/\veps_{\textbf{k}}
$
and
$
	1/\tautr(\textbf{k})
	= 
	(m \pi \lamimp/ v_z)
	\veps_{\textbf{k}}
	.
$ Physically, $\omega_{c,\sfeff}$ represents the effective cyclotron frequency of electrons. 
Interestingly, $\omega_{c,\sfeff}$ flips sign across the nodal ring.
Meanwhile, $1/[2\tautr(\textbf{k})]$ represents the impurity scattering rate for transport.
For small $T$, $\veps_{\textbf{k}}$ is pinned at the chemical potential $\mu$ and thus the transport rate is approximately a constant. }

\begin{figure}
\centering
\includegraphics[width=1\linewidth]{{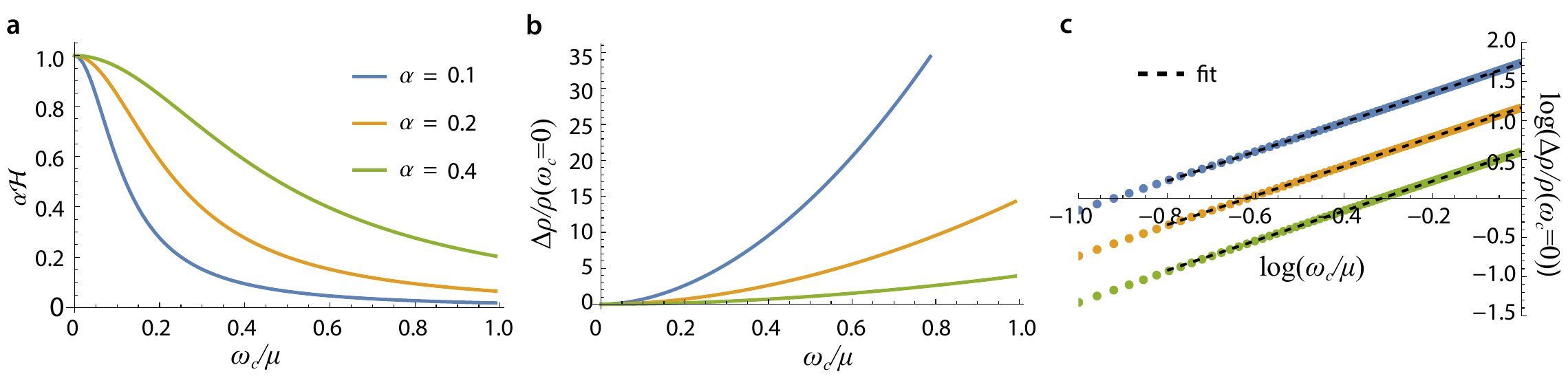}}
\caption{\textbf{a} $\calH(\omega_c,\mu,\alpha)$ as a function of field at three different impurity strengths. Note that in the plot the field is in a normalized unit, $\omega_c/\mu$. Here $\alpha$ is a parameter controlling the impurity strength, $\omega_c$ is the cyclotron frequency and $\mu > 0$ is the chemical potential. \textbf{b} $\Delta \rho_{xx}/\rho_{xx}$ as a function of field.  \textbf{c} Same as (b), but in $\log-\log$ scale. The linear fit (black dashed lines) reveals $\Delta \rho_{xx}/\rho_{xx} \sim \omega_c^{\beta}$, with $\beta \simeq (1.87,1.88,1.85)$ for $\alpha = (0.1,0.2,0.4)$ respectively. In all these panels, we have set $\mu = 1$.}
\label{fig:plot_H_combine}
\end{figure}

\textcolor{black}{For $\mu > 0$ and $T \rightarrow 0$, the current along $\textbf{E} \parallel \hat{x}$ is solely contributed by electrons,
\begin{align}
	J^x_e
	=
	-e
	\int\limits_{\textbf{k}}
	v_{\textbf{k}}^x
	\,
	\delta f_e(\textbf{k})
	=
	\sigma_{xx}(\omega_c) E,
\end{align}
where
\begin{eqnarray}
	\sigma_{xx}(\omega_c)
	=
	\sigma_0 \, \calH(\omega_c,\mu,\alpha),
	,
	\qquad
	\calH(\omega_c,\mu,\alpha)
	=
	2\, \alpha \, \frac{\mu^2}{\omega_c^2}
	\left(
		1 - \frac{\alpha}{\sqrt{\alpha^2 + (\omega_c/\mu)^2}}
	\right).
\end{eqnarray}
Here $\alpha = m\pi \lamimp/v_z$ is a dimensionless quantity characterizing the disorder strength and $\sigma_0 = \frac{e^2 E_M}{4\pi v_z}$ has the dimension of conductivity.
Note that for $\omega_c \rightarrow 0$, ${\cal H}(\omega_c \rightarrow 0,\mu,\alpha) \rightarrow \alpha^{-1}$ and we recover the conductivity at $B = 0$: $\sigma_{xx}(\omega_c = 0) = \sigma_0/\alpha$. 
The $\omega_c$ dependence of the function ${\cal H}$ is shown in Fig.~\ref{fig:plot_H_combine}a.}

\textcolor{black}{Meanwhile, the Hall current is written as:
\begin{align}
	J^y_e
	=
	-e
	\int\limits_{\textbf{k}}
	v_{\textbf{k}}^y
	\,
	\delta f_e(\textbf{k})
	=
	\label{eq:sigma_xy_KT}
	e^2 E
	\int\limits_{\textbf{k}}
	\left(
		- \frac{\partial f_{e,0}}{\partial \veps_{\textbf{k}}}
	\right)
	\left(
		\frac{\xikpara}{\veps_{\textbf{k}}}
		\frac{k_{\parallel}}{m}
		\sin \phi
	\right)^2
	\tautr(\textbf{k})
	\frac{
		\omega_{c,\sfeff}(\textbf{k}) \tautr(\textbf{k})
	}{
		1 + [\omega_{c,\sfeff}(\textbf{k}) \tautr(\textbf{k})]^2
	},
\end{align}
which is approximately $0$ for large $E_M$, because $\omega_{c,\sfeff}(\textbf{k})$ flips sign across the nodal ring as the velocity $\textbf{v}_{\textbf{k}}$ is reversed, implying that
$
\sigma_{xy} \simeq 0.
$
As we show below, the negligible Hall conductivity has important consequences for the nonsaturating MR observed in a Weyl nodal-ring system.}

\textcolor{black}{Recall that in a normal one-band metal, the conductivity tensor is given by [\textcolor{blue}{35}]
\begin{align}
	\hsig^{\sfmetal}
	=
	\begin{bmatrix}
	\sigma_{xx}^{\sfmetal} & \sigma_{xy}^{\sfmetal}
	\\
	-\sigma_{xy}^{\sfmetal} & \sigma_{xx}^{\sfmetal}
	\end{bmatrix},
	\end{align}
where
\begin{eqnarray}
	\qquad
	\sigma_{xx}^{\sfmetal} 
	=
	\frac{1}{1 + (\omega_c \tauel)^2}\sigma_{\sfDrude} ,
	\qquad
	\sigma_{xy}^{\sfmetal} 
	=
	\frac{\omega_c \tauel}{1 + (\omega_c \tauel)^2} \sigma_{\sfDrude} ,
\end{eqnarray}
In the equations, $\sigma_{\sfDrude}$ is the Drude conductivity and $\tauel$ is the elastic scattering rate due to disorder. 
The resistivity tensor is obtained by matrix inverse operation: $\hat{\rho}^{\sfmetal} = [\hsig^{\sfmetal}]^{-1}$. It leads to $\rho_{xx}^{\sfmetal} = \sigma_{\sfDrude}^{-1}$, which has no $B$ field dependence at all. 
However in Weyl nodal ring semimetal systems, the velocity sign flip leads to neglibily small Hall conductivity ($\sigma_{xy} \simeq 0$). Thus the transverse conductivity can be simply obtained by $\rho_{xx} \simeq 1/\sigma_{xx}$, implying a \textit{nonsaturating} MR behavior which does not require perfect compensation of electrons and holes. }

\textcolor{black}{In Fig.~\ref{fig:plot_H_combine}b, we plot $\Delta \rho_{xx}/\rho_{xx}= 1/(\alpha \calH) - 1$ as a function of $\omega_c/\mu$ for various impurity strength controlled by $\alpha$. 
The nonsaturating behavior of $\rho_{xx}$ is clear. 
The field dependence of $\Delta \rho_{xx}/\rho_{xx}$ can be fitted well with a power function, i.e. $\Delta \rho_{xx}/\rho_{xx} \sim \omega_c^{\beta}$, as shown in the log-log plot in Fig.~\ref{fig:plot_H_combine}c. 
The linear fittings reveal that the power $\beta$ is approximately in the range between $1.8$ and $1.9$.} 

\subsection{B. Quantum theory for magnetoconductivity}
\hfill \break
\textcolor{black}{While the semiclassical theory can well describe the low-field MR behavior, its accuracy is undermined when the system enters the Landau quantized regime, especially when a large enough field is applied so that the system approaches the quantum limit. We now turn into a fully quantum mechanical description of the MR using the Kubo formula. }

\textcolor{black}{As a first step, we calculate the Landau level spectrum. 
With the magnetic field $\textbf{B} = B \hat{z}$, we employ the gauge
$
	A_y = B x, \; A_x = A_z = 0
$, 
where $\textbf{A}$ is the vector potential, 
and send
$
	k_y \rightarrow k_y + e A_y
$ 
in the Hamiltonian in Eq.~(\ref{eq:hamiltonian}). 
We also account for the intrinsic non-flatness (energy variation) of the nodal ring by introducing an extra term 
$
	\delta \hh(\textbf{k})
	=
	\lambda k_x^2/2m
$
into the original model Hamiltonian in Eq.~(\ref{eq:hamiltonian}). 
Here, $0 < \lambda < 1$ is a small parameter controlling the level of energy variation of the nodal ring. 
We treat such variation perturbatively for small $\lambda$. 
The eigenenergies at Landau level $n$ is given by
\begin{eqnarray}
	\pm E_{n,k_z} 
	=
	\pm \sqrt{
		(\veps_n - E_M)^2 + (v_z k_z)^2
	}
	+
	\delta E_n
	,
	\qquad
	\veps_n 
	=
	\left(n + \frac{1}{2}\right)\omega_c,
	\qquad
	\delta E_n
	\simeq
	\lambda 
	\omega_c
	\frac{2 n + 1}{4},
\end{eqnarray}
where $\delta E_n$ is the energy shift due to $\delta \hh$ at order ${\cal O}(\lambda)$. 
The corresponding eigenstates can be expressed in terms of the Hermite polynomials.}

\textcolor{black}{We compute the $xx$ and $xy$ components of the current-current correlation function following the standard procedures outlined in Refs.~[\textcolor{blue}{36}, \textcolor{blue}{37}]. 
In the presence of disorder, we assume that the eigenstates are approximately unchanged and introduce a constant self-energy $\Gamma_B$, describing phenomenologically the impurity scattering rate as inspired by the kinetic theory results. 
We confirm that our expressions reduce to the ones based on the kinetic theory in the semiclassical limit for small field $B$ and weak disorder. 
Physically, the Kubo calculation accounts for the discreteness of the Landau levels and the smearing of the spectral function, in addition to the semiclassical motion of electrons.} 

\begin{figure}[ht]
\centering
\includegraphics[width=0.72\linewidth]{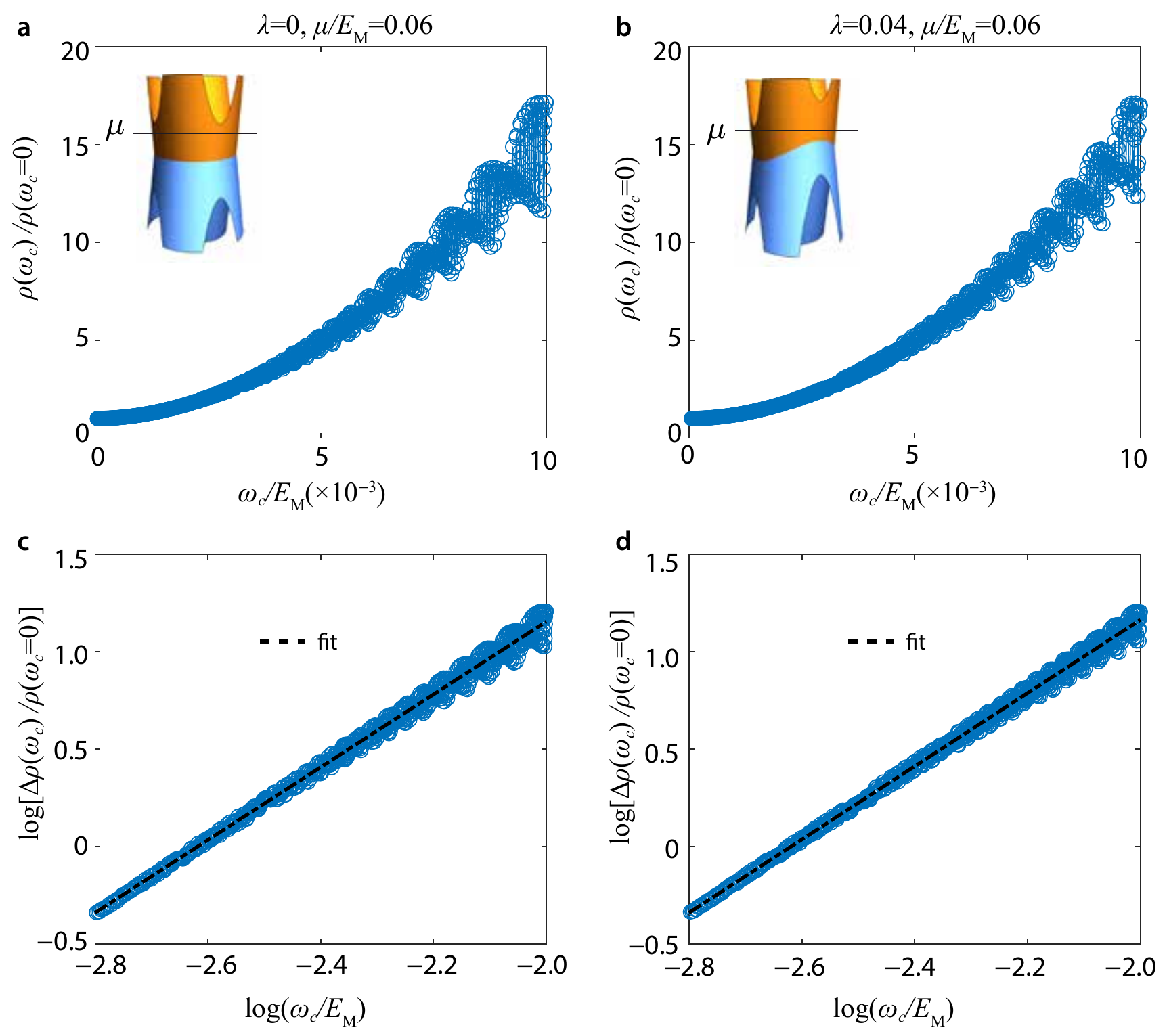}
\caption{
Field dependent resistivity normalized by its zero-field value for (\textbf{a}) $\lambda = 0$ (flat nodal ring), (\textbf{b}) $\lambda = 0.04$ (non-flat nodal ring). In both cases, we set the chemical potential $\mu = 0.06E_M$ and impurity scattering rate $\Gamma_B = 10^{-3}E_M$. 
The insets show the corresponding band structure and chemical potential at $\omega_c = 0$ and $k_z = 0$. 
Note that for $\lambda > 0$, the energy of the nodal ring is shifted upward by an amount $\delta E / E_M \simeq \lambda/2$. 
\textbf{c,d} The corresponding log-log plots.
The linear fit (black dashed lines) for the $\omega_c/E_M > 1.5 \times 10^{-3}$ data reveals $\Delta \rho_{xx}/\rho_{xx} \sim \omega_c^{\beta}$, with $\beta \simeq 1.90$ for both $\lambda = 0$ and $0.04)$. 
}
\label{fig:MR_Kubo_normal_and_lnln}
\end{figure}

\textcolor{black}{Below we present the field dependence of the MR by numerically evaluating the Kubo expression. 
We now consider a practical $E_M$ value with the energy scale comparable to that of the Weyl nodal rings that lead to the formation of the $\beta$ pockets in EuGa$_4$ (Fig.~\ref{figureS5}b), and a small $\lambda$ so that the Fermi surface in this model forms a torus geometry. Under this condition, we have $\sigma_{xy} \ll \sigma_{xx}$ and  $\rho_{xx} \simeq 1/\sigma_{xx}$.  
In the numerical calculation, we summed over $n_{\sfmax} = 50000$ Landau levels to ensure convergence. 
The results for a representative set of parameters are shown in Fig.~\ref{fig:MR_Kubo_normal_and_lnln}. Note that the \textit{x}-axis in the figures is represented by a normalized field parameter, $\omega_c/E_M$.}

\textcolor{black}{We fix the chemical potential to be a positive value of $\mu = 0.06 E_M$ and the impurity scattering rate to be $\Gamma_B = 10^{-3} E_M$. 
The small scattering rate is justified given the high carrier mobility of the sample. 
We assume the field dependence of $\mu$ and $\Gamma_B$ is weak and thus negligible. 
This is approximately valid at a finite $\mu > \omega_c$. 
The $\mu \rightarrow 0$ scenario is interesting but requires special attention [\textcolor{blue}{36}, \textcolor{blue}{38}, \textcolor{blue}{39}]. We will leave it for future studies. } 

\textcolor{black}{For the case of flat Weyl nodal ring $(\lambda = 0$) (Fig.~\ref{fig:MR_Kubo_normal_and_lnln}a), $\rho_{xx}$ shows an approximately quadratic field dependence at low fields ($\omega_c/E_M<\sim1\times 10^{-3}$) and does not exhibit
obvious quantum oscillations, in agreement with the results from the kinetic theory. 
As the field further increases, the system enters the Landau quantized regime, and the resistivity demonstrates gradually enhanced oscillations due to the discreteness of the Landau level energy spectrum. With the existence of the inner and outer extremal cyclotron orbits of the nodal ring, the oscillations also show a beating pattern.} 

\textcolor{black}{For the case of non-flat Weyl nodal ring $(\lambda > 0$), the results are shown in 
Fig.~\ref{fig:MR_Kubo_normal_and_lnln}b. 
Here we assumed a relatively small energy variation, $\lambda = 0.04$. In this case, $\mu$ does not cross the nodal ring at $k_z = 0$ in the zero field limit (see the illustration in the inset). This scenario bears resemblance to the nodal rings that lead to the formation of $\beta$-pockets in EuGa$_4$. The field dependence is qualitatively the same as the $\lambda = 0$ case (Fig.~\ref{fig:MR_Kubo_normal_and_lnln}a) except the change in the oscillation frequencies.}

\textcolor{black}{In both cases, we show the numerical calculations up to the field, $\omega_c/E_M=0.01$, which is about $1/100$ of the field strength that is required for the system to reach quantum limit. This is comparable to the applied field strength in our measurements: the maximal measured field of $\sim 40$~T is about $1/100$ of the field strength that is required for the electrons with the $\beta$-pockets to reach quantum limit.} 


\textcolor{black}{In Figs.~\ref{fig:MR_Kubo_normal_and_lnln}c,d, we show the log-log plots of the MR curves. The linear fits of the high-field data (black dashed lines) indicate that the MR follows the power function $\Delta \rho_{xx}/\rho_{xx} \sim \omega_c^{\beta}$, with the exponent $\beta \simeq 1.90$ for both flat and nonflat Weyl nodal rings.} 

\subsection{C. Discussion}
\begin{figure}[ht]
\centering
\includegraphics[width=0.5\linewidth]{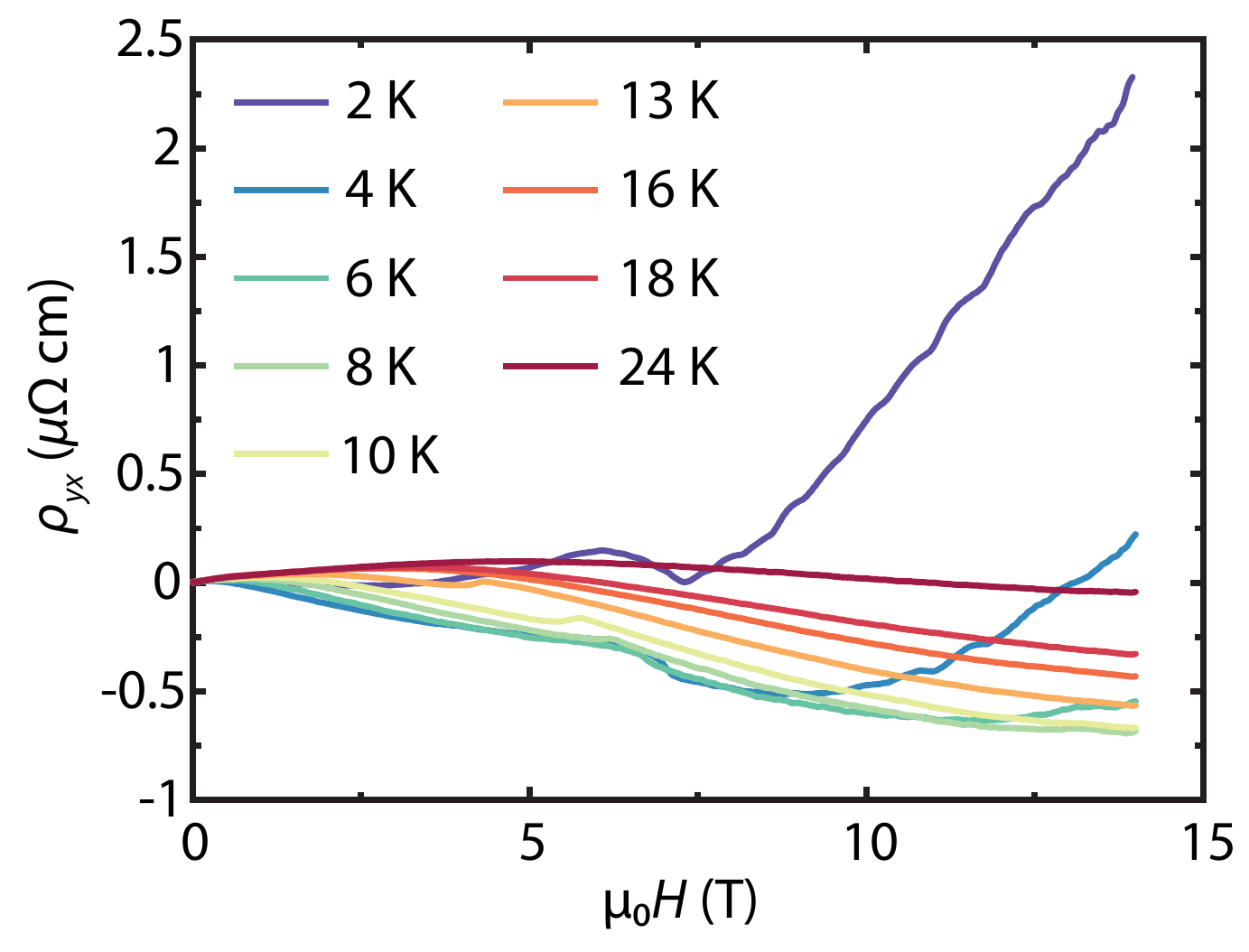}
\caption{
Hall data measured at a series of temperatures from 2~K to 24~K. 
}
\label{Hall}
\end{figure}

\hfill \break
\textcolor{black}{We showed above the magnetotransport results based on semiclassical and quantum theory for a Weyl nodal ring system. we find that the nonsaturating MR naturally arises in a Weyl NR system, without the stringent requirement of perfect electron-hole carrier compensation [\textcolor{blue}{17}, \textcolor{blue}{40}]. In fact, we only assumed one type of conducting carriers in our theoretical model. This unusual behavior benefits from the negligibly small Hall conductivity, which derives from the sign reversal of the Fermi velocity across the nodal ring. We performed the Hall measurements on the high-quality EuGa$_4$ single crystal, and show the data in Fig.~\ref{Hall}. The Hall resistivity, $\rho_{yx}$, is indeed significantly smaller than the transverse resistivity, $\rho_{xx}$ (see Fig.~\ref{figureS10}). At 2~K and 14~T, $\rho_{yx}/\rho_{xx}$ is only about $2\%$, which supports the treatment of $\rho_{xx}\simeq1/\sigma_{xx}$ in our magnetotransport model.}

\textcolor{black}{We note that in our model, we did not consider the contribution of the small $\gamma-$pockets. Apparently, as $\mu \rightarrow 0$, electrons of the $\gamma-$pockets can be forced to occupy a few discrete Landau levels at much lower field than that of the bigger pockets. The magnetotransport properties in this scenario are interesting, but require special treatment [\textcolor{blue}{36}, \textcolor{blue}{38}, \textcolor{blue}{39}], as we mentioned above. Nevertheless, it is safe to conclude that the contribution of the 
$\gamma-$pockets to the nonsaturating MR behavior is small, given the small fraction of the carriers from these pockets. This is in sharp contrast to the quantum magnetoresistance mechanism proposed by Abrikosov [\textcolor{blue}{36}], where nonsaturating MR appears only when electrons occupy the lowest Landau level. Therefore, we consider Weyl nodal ring semimetals as a novel platform to host the nonsaturating MR.}

\section{\label{sec13}Structural refinement from powder X-ray diffraction}
\begin{figure*}[htb]
    \centering
  \includegraphics[width=0.5\textwidth]{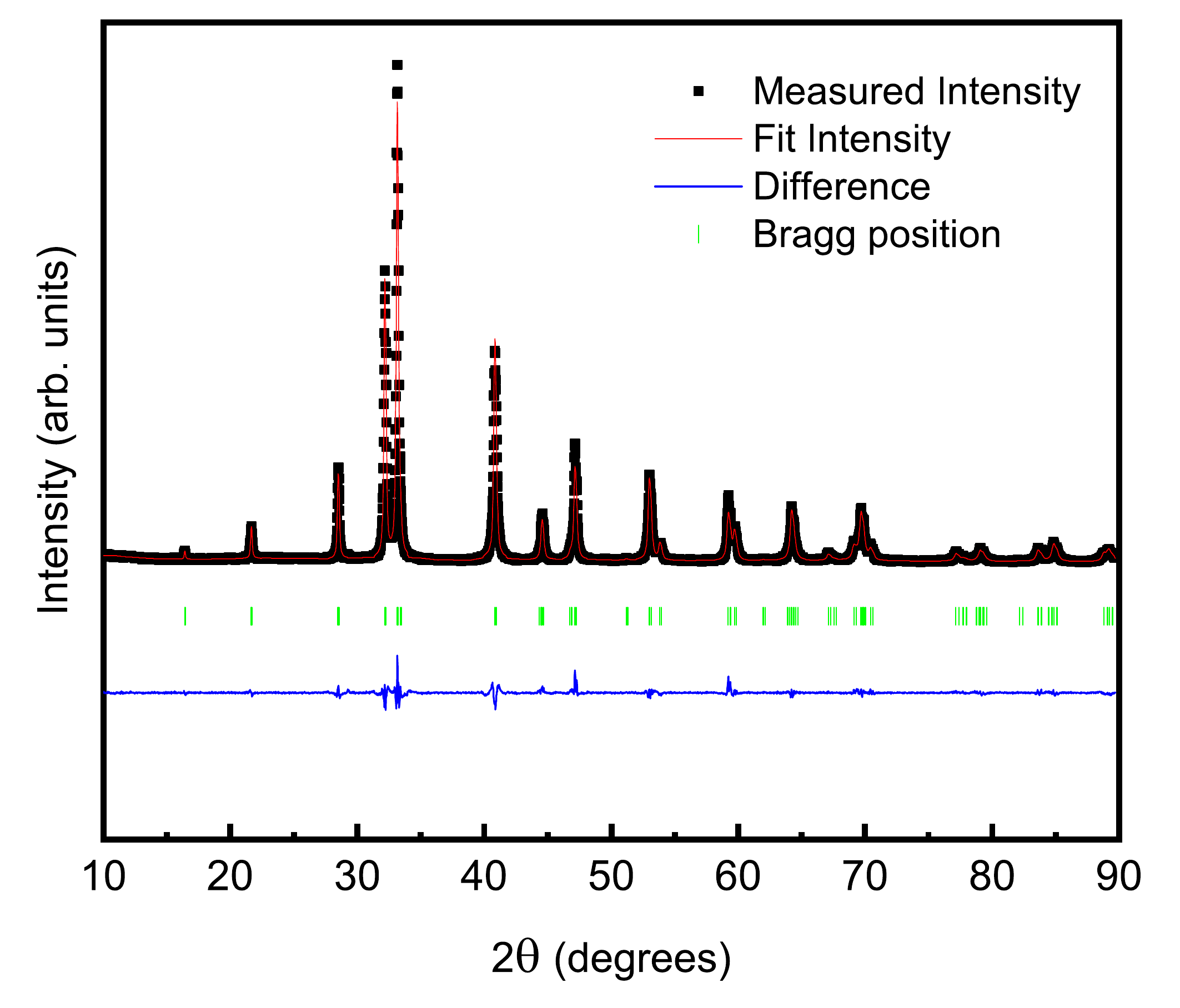}
\caption{Powder x-ray diffraction pattern of EuGa$_{4}$ taken at T = $300\,$K along with the Rietveld refinement (red line), the blue line is the difference between the measured and the fitted intensity while the green ticks correspond to the Bragg peak positions. 
}
  \label{figureS15}
\end{figure*}

Table ~\ref{structureparameter} provides the atomic positions for the structure of EuGa$_{4}$ from Rietveld refinement along with the corresponding lattice parameters. We carried out powder x-ray diffraction measurements and the corresponding diffraction peaks are shown in Fig. \ref{figureS15}.

\begin{center}
\begin{table}[htp]
\caption{Structural parameters for EuGa$_{4}$ at 300K. Space group I4/mmm (No. 139).}
\vspace{3mm}
    \centering
\begin{tabular}{ >{\centering\arraybackslash}m{1.5cm}|>{\centering\arraybackslash}m{1.5cm}|>{\centering\arraybackslash}m{2cm}|>{\centering\arraybackslash}m{2cm}|>{\centering\arraybackslash}m{2cm}|>{\centering\arraybackslash}m{2cm} } 
 \hline
 Atom & Wyckoff & Occupancy & x & y & z\\ 
 \hline
 Eu & 2a & 1 & 0 & 0 & 0\\ 
 Ga1 & 4e & 1 & 0 & 0 & 0.38388(17)\\ 
 Ga2 & 4d & 1 & 0 & 0.5 & 0.25\\ 
 \hline
\end{tabular}
\begin{center}
\begin{tabular}{>{\centering\arraybackslash}m{11.85cm}}
a = 4.39564(7) \AA, c = 10.66121(19) \AA \\
R$_{wp}$=11.43$\%$, R$_{exp}$ = 7.04$\%$\\
\hline
\end{tabular}

\end{center}
\label{structureparameter}
\end{table}

\end{center}

%
\newpage


\end{spacing}